%% file: pap3_main.tex
\documentclass[prc,showpacs,floatfix]{revtex4}
\usepackage{bm}% bold math
\usepackage{dcolumn}
\usepackage{graphicx}% Include figure files
\begin{document}
%\preprint{MKPH-T-05-1}
\title{
\hfill{\small {\bf MKPH-T-05-1}}\\
{\bf Double pion photoproduction on nucleon and deuteron}}
\author{A. Fix and H. Arenh\"ovel}
\affiliation{
Institut f\"ur Kernphysik,
Johannes Gutenberg-Universit\"at Mainz, D-55099 Mainz, Germany}
\date{\today}
\begin{abstract}
Photoproduction of two pions on nucleon and deuteron is studied for
photon energies from threshold up to $E_\gamma = 1.5$~GeV. For the
elementary operator an effective Lagrangean approach is used with
resonance and Born contributions. The model parameters are fixed by
resonance decay widths and multipole analyses of single pion
photoproduction. A satisfactory description of total cross sections of
two-pion production on the proton for various charge channels is
achieved, except for $\pi^0\pi^0$ production for which a significant
underestimation is found near threshold. The operator then
serves for the evaluation of this reaction on the deuteron in the
impulse approximation. In addition, $NN$ rescattering in the final
state is taken into account, but $\pi N$ and $\pi\pi$ rescatterings
are neglected. Results are presented for total cross sections and
target asymmetries. 
\end{abstract}

\maketitle

%\pacs{PACS numbers:  13.60.Le, 21.45.+v, 25.20.Lj}

\input{pap3_ch1}

\input{pap3_ch2}

\input{pap3_ch3}
\input{pap3_ch4}

\input{pap3_app}

\input{pap3_ref}
\newpage
\input{pap3_tab}
\newpage
\input{pap3_fig}

\end{document}

%% file: pap3_ch1.tex
\section{Introduction}

Double pion photoproduction is another important tool for our 
understanding of nucleon structure besides single pion production. It is 
usually considered as a complementary reaction serving as the main
source of information which can not be obtained otherwise, e.g.\ from 
single pion photoproduction. In particular, this process is quite promising
for the  study of the so-called ``missing'' resonances which are only weakly 
coupled to the $\pi N$ channel~\cite{Capstic}. 

The elementary reaction $\gamma N\to \pi\pi N$ is clearly much more complicated
than single pion photoproduction. This fact is reflected in the existing
theoretical approaches which show a strong model dependence of the
results~\cite{Oset,Laget,Ochi,Mokeev}. Among the reasons for this model 
dependence, we firstly would like to note that the Born amplitude,
which is considerably more important than it is in single pion production, 
has a very complicated structure. Another reason lies in the presence of 
various interactions
between the final particles (FSI). This problem is not present in 
elementary single pion photoproduction because $\pi N$ rescattering is
already included in the elementary amplitude. So far the
most advanced investigation of FSI was performed within the WKB
model~\cite{Luke} using as an essential ingredient the dominance of
meson exchange in the production mechanism at high energies. It
produces quite a strong absorptive effect, which changes dramatically
the energy dependence of the cross section. An essentially different
approach was realized in~\cite{Mokeev} where the $\pi NN$ interaction
was reduced to resonance excitations in the quasi two-body $\pi\Delta$
and $\rho N$ systems. In contrast to the WKB results, the last method
predicts quite an insignificant role of FSI. These difficulties are
among others the main reason why the mechanisms of the elementary double pion
photoproduction so far appear only partially understood. 

The analysis of the various theoretical approaches is usually focused on the
experimentally favorable case where the target is a proton, whereas the
results on the neutron depend on the model assumptions used for extracting the
data from measurements on the deuteron or on other light nuclei.  
The neutron data are obviously needed for a systematic analysis of the
isotopic spin structure of the elementary amplitude. The main
question arising in this connection is, what is the role of the
``nuclear effects'', e.g., Fermi motion, final state interaction, and
two-nucleon production contributions, which prevent a model 
independent study of the neutron amplitude. Whereas the Fermi motion
is naturally included in the spectator model, the interaction between
the final particles requires considerably more calculational efforts.

Corrections to the mere quasi free production were partially
considered in~\cite{Zabrod}. It was shown that the experimental yields
for $\pi^+\pi^-$ photoproduction were almost the same for 
$^1$H and $^2$H targets, so that the total effect from higher order
processes in the latter case was expected to be insignificant. 
The strong validity of the spectator model was also assumed in~\cite{Kleber} 
for the extraction of the $\gamma n\to\pi^0\pi^0n$ cross section. In
fact, it is often concluded that FSI and other
higher order processes have an insignificant influence simply on the
grounds of a good agreement between the data and the impulse
approximation. However, such a conclusion is not stringent and might
be misleading. Therefore, it is the aim of the present work to study
the role of FSI using a refined phenomenological model for the 
basic elementary amplitude and including in addition the final $NN$
interaction. 

Accordingly, the paper is divided into 
two parts. The first one is devoted to the elementary reaction while
the second deals with purely nuclear effects in the reaction on a
deuteron. In Sect.~2 we describe briefly our $\gamma N\to \pi\pi N$ model. 
The emphasis lies on those points, where our approach differs from 
previous work \cite{Oset,Laget,Ochi,Mokeev}.
The $\pi\pi$ photoproduction on the deuteron is considered
in Sect.~3 where the results are presented and discussed. 
In several appendices we describe in detail the formal
ingredients of the elementary production operator.

%% file: pap3_ch2.tex
\section{The $\gamma N\to \pi\pi N$ model}\label{gammapipi}

In this section we will outline the formalism for the photoproduction of
two pions on the nucleon 
\begin{equation}\label{0}
\gamma(k,\vec{\epsilon}_\lambda)
+N(p_i)\rightarrow \pi^{\mu_1}(q_1)+\pi^{\mu_1}(q_2)+N(p_f) \,,
\end{equation}
where the 4-momenta of the participating particles, incoming photon,
initial and final nucleon and the two pions, are respectively denoted by
\begin{equation}\label{5}
k=(\omega_\gamma,\vec{k}),\, p_{i/f}=(E_i,\vec{p}_{i/f}),\,
q_{1/2}=(\omega_{1/2},\vec{q}_{1/2})\,.
\end{equation}
The circular polarization vector of the photon is described by 
$\vec{\epsilon}_\lambda$ with $\lambda=\pm 1$ and the superscript 
$\mu_{i}=0,\pm 1$ in (\ref{0}) denotes the charge of the $i$th pion. 

Using standard covariant normalization for the free particle states
\cite{Bjork}, the unpolarized differential cross section in the
overall c.m.\ frame can be expressed in terms of the reaction matrix
$t^{\mu_1\mu_2}_\lambda$ 
\begin{equation}\label{10}
d\sigma
=(2\pi)^{-5}\frac{M_N^2}{8W^2}\ \frac{q^*p}{\omega_\gamma}\, \frac{1}{4}
\sum\limits_{\mbox{\tiny spins}} |t^{\mu_1\mu_2}_\lambda
(\vec q_1,\vec q_2,\vec k\,)|^2\ dw_{\pi\pi}
d\Omega_{q^*}d\Omega_p\,,
\end{equation}
where $W$ denotes the total c.m.\ energy, $\vec p=-(\vec q_1+\vec q_2)$ 
the final nucleon momentum, 
and the variables $w_{\pi\pi}$ and $\vec q^{\,\ast}$ stand for the invariant
$\pi\pi$ mass and the relative momentum in the c.m.\ system of the two
pions, respectively. The spin structure of the reaction amplitude can
be expressed in terms of the nucleon Pauli spin matrices as 
\begin{equation}\label{15}
t^{\mu_1\mu_2}_\lambda(\vec q_1,\vec q_2,\vec k\,)=
K^{\mu_1\mu_2}_\lambda(\vec q_1,\vec q_2,\vec k\,)
+i\,\vec{L}^{\mu_1\mu_2}_\lambda(\vec q_1,\vec q_2,\vec k\,)
\cdot\vec{\sigma}\,.
\end{equation}
Since two pseudoscalar mesons are produced, $K^{\mu_1\mu_2}$ must be a
scalar and $\vec{L}^{\mu_1\mu_2}$ a pseudovector. 

The corresponding isospin decomposition reads, as is discussed in
detail in Appendix~\ref{AppA} (see Eq.~(\ref{isospinstructure})),
\begin{eqnarray}\label{20}
t^{\mu_1\mu_2}(\vec q_1,\vec q_2,\vec k\,)&=&\sum_{l=1}^3 
\Big(f_l^{(+)}(\vec q_1,\vec q_2,\vec k\,){\cal O}_{\mu_1\mu_2}^{l(+)}
+f_l^{(-)}(\vec q_1,\vec q_2,\vec k\,){\cal O}_{\mu_1\mu_2}^{l(-)}\Big)\,,
\end{eqnarray}
where the operators ${\cal O}_{\mu_1\mu_2}^{l(\pm)}$ are defined in
Eqs.~(\ref{A5}) through (\ref{iso_O3pm}). The spin-spatial functions
$f_l^{(\pm)}$ depend only on the pion momenta and are independent of
their charges $\mu_1$ and $\mu_2$. They have an analogous spin
structure as in (\ref{15}). The functions $f_l^{(+)}$ are symmetric
with respect to the two pion momenta and contribute to two-pion states
with total isospin $T=0,2$, and the $f_l^{(-)}$ are antisymmetric and
contribute to $T=1$ states only. 

Our calculation of the reaction amplitude $t^{\mu_1\mu_2}$ follows the
same lines as in \cite{Oset,Laget,Ochi,Ong}. Namely, we use the
traditional phenomenological Lagrangean approach with Born and
resonance contributions on the tree level. Multiple scatterings within
the $\pi N$ and $\pi\pi$ subsystems are effectively taken into account
by introducing nucleon and meson resonances, respectively. For the
resonance contributions, the final two-pion state then results from a
two-step decay via intermediate quasi-two-body channels for which we
take here $\pi \Delta$, $\rho N$ and $\sigma N$ channels. Thus the
corresponding amplitude can be presented schematically in the form 
\begin{equation}\label{25}
{t}={t}^B+{t}^{\pi\Delta}+{t}^{\rho N}+{t}^{\sigma N}\,,
\end{equation}
where $t^B$ contains all Born terms. The specific diagrams used in the
calculation are presented in Fig.~\ref{fig1}. Since the operator
(\ref{25}) will be implemented into the deuteron, the corresponding
amplitudes are treated non-relativistically with respect to the
baryons keeping only terms up to the order $(p/M_N)$, denoting the
nucleon mass by $M_N$. 

The resonances included in the model are those which are localized in
the mass region up to 1.8~GeV and classified with four stars in the
Particle Data Group compilation~\cite{PDG}. Only the $S_{11}(1650)$
was ignored because of its insignificance. All resonances are listed
in Table~\ref{tab1}. Since we want to consider the reaction up to
energies of $E_\gamma = 1.5$~GeV, resonances with higher spin $J=5/2$
and both parities were included. The hadronic coupling constants were
fitted to the corresponding decay widths taken from~\cite{PDG}. Their
values are listed in Tables~\ref{tab3} and \ref{tab3a}. In the
quasi-two-body decays $N^*\to \pi\Delta$, $N^*\to \rho N$, and $N^*\to
\sigma N$, the finite widths of $\Delta$, $\rho$, and $\sigma$ were
taken into account. As independent parameters, characterizing the
electromagnetic transitions $\gamma N\to N^*$, we used the helicity
amplitudes also taken from~\cite{PDG}. More details of the formalism
are presented in several appendices. The signs of the $\pi N$
constants are chosen in such a manner that the multipole amplitudes
for pion photoproduction, obtained with our Lagrangeans, are
consistent with the corresponding amplitudes of the standard multipole
analyses (see e.g.~\cite{MAID}). The choice of the sign for the
$N^*\to\pi\Delta$ and $N^*\to\mu N$ ($\mu\in\{\rho,\sigma\}$) decay
amplitudes was based on the $\pi N\to\pi\pi N$ analysis
of~\cite{Manley}. 

The next point which deserves a comment is an absorptive effect which
is quite well known from photoproduction of vector mesons at high
energies. As is discussed in Ref.~\cite{Gottfried}, the absorption
follows simply from the existence of many inelastic channels which
compete with the process under consideration so that the resulting
cross section must essentially be lower than that predicted by the
Born approximation alone. Here we follow the prescription
of~\cite{Locher} which was used also in~\cite{Luke,Laget}. Namely, the
$\Delta$ Kroll-Ruderman term (see diagram (8) of Fig.~\ref{fig1}) was
multiplied by an energy dependent attenuation factor
\begin{equation}\label{30} 
\eta=\Big(1-Ce^{-(J-1/2)A q^2/2}\Big)^{1/2}\,, 
\end{equation}
with $J=3/2$ as the total $\pi\Delta$ angular momentum. The
$\pi\Delta$ c.m.\ momentum is denoted by $q$, and the parameter values
$C=1$ and $A=8$ GeV$^{-2}$ were taken from~\cite{Luke}. The same
prescription was employed for $\Delta$ exchange in the $u$-channel
(second diagram of (14) of Fig.~\ref{fig1}) where we assume a weak
angular dependence of the $\Delta$ $u$-pole propagator, at least at
forward angles. For the $\rho$ Kroll-Ruderman term (diagram (4)) we
have taken for the attenuation factor $C=1$ and
$A=5.5$~GeV$^{-2}$~\cite{Luke}. In this case $q$ denotes the
$\rho$-meson c.m.\ momentum. As one can see from (\ref{30}) with
$C=1$, the $J=1/2$ part vanishes completely because of absorption, and
the resulting contribution from charged $\rho$-photoproduction
(Fig.~\ref{fig2a}) turns out to be relatively small. 

For those diagrams containing a meson exchange in the $t$ channel
(diagrams (2), (5)-(7), (9), and (10) of Fig.~\ref{fig1}) we adopt the
sharp cutoff approximation of~\cite{Dar}. It is based on the
assumption that the final particles are completely absorbed within a
sphere of radius $R$. Then the dependence of the amplitude on the
invariant Mandelstam variable $t$ is changed as follows 
\begin{eqnarray} \label{35}
\frac{1}{t-\mu^2}&=&-\int\limits_0^\infty bdb\, J_0(b\sqrt{-t})K_0(b\mu)
\nonumber \\
&\rightarrow& -\int\limits_R^\infty bdb\, J_0(b\sqrt{-t})K_0(b\mu)=
\frac{R}{t-\mu^2}[\mu J_0(R\sqrt{-t})K_1(R\mu)-\sqrt{-t}
J_1(R\sqrt{-t})K_0(R\mu)]\,,
\end{eqnarray}
where $J_n$ and $K_n$ are cylindrical and hyperbolic Bessel 
functions~\cite{AbSt},
and $\mu$ denotes the mass of the exchanged meson. The integration
variable $b$ in (\ref{35}) can be interpreted as an impact
parameter. The radius $R$ of the absorbing sphere was chosen such that
the characteristic decrease of the experimental cross section for
$\pi^+\pi^-$ photoproduction on a proton above $E_\gamma=1$~GeV is
reproduced. We have taken $R=0.15$~fm for all three exchanged mesons
$\pi$, $\rho$ and $\sigma$. Due to this absorptive effect, the central
part of the final state wave function vanishes leading to a sharpening
of the peripheral peak in the angular distribution as is demonstrated
in Fig.~\ref{fig3}. 

As already stressed in the introduction, the quasi-classical nature of
this method makes its application doubtful in the second resonance
region, where the $\pi\Delta$ and $\rho N$ interactions are not of
diffractive character, and other aspects of the final state
interaction should come into play. Thus we consider this method only
as a simple, heuristic possibility to reduce a too strong increase of
the cross section above $E_\gamma$=750 MeV, which on the other hand
can be physically motivated, in contrast to fitting the data with
extremely soft form factors. 

Our results for the total cross section of two-pion production on the
proton are presented in Figs.~\ref{fig2} through \ref{fig2b}. The pion
photoelectric term (diagram (9) of Fig.~\ref{fig1}) is well known to
give most of the forward charged $\pi$ production in the $\pi \Delta$
channel. The $\Delta$ Kroll-Ruderman term, needed for restoring gauge
invariance of the amplitude, provides an essential part of the total
cross section for $\pi^+\pi^-$ production. With respect to other Born
diagrams, as is shown in the lower panel of Fig.~\ref{fig2}, 
important contributions come also from the nucleon Kroll-Ruderman and
pion-pole terms (diagrams (1) and (2) of Fig.~\ref{fig1}) as well as
from the $N\Delta$ and $\Delta\Delta$ $s$- and $u$-channels (diagrams
(12) and (14) of Fig.~\ref{fig1}). Also $\sigma$ exchange in
$\rho$-photoproduction (diagram (6)) is responsible for quite a large
fraction of the $\pi^+\pi^-$ cross section above $E_\gamma=1.2$~GeV.
The remaining Born terms are less important, but their combined effect
becomes still significant with increasing energy (long dash-dotted
curve in the lower panel of Fig.~\ref{fig2}). 

In order to compare our results with those of other authors, we
present in Figs.~\ref{fig2} through \ref{fig2b} separately the
contributions of the individual resonances to the total cross
section. The comparison to previous work shows that our model is quite
close to that of the Valencia group~\cite{Oset} but differs visibly
from the Saclay model~\cite{Laget} which, however, is primarily
concerned with the role of the Roper resonance. The only essential
difference to~\cite{Oset} lies in the treatment of the $D_{13}(1520)$ 
resonance. The spin structure of the $D_{13}\to\pi\Delta$ transition
used in~\cite{Oset} leads to an additional strong momentum dependence
in the $s$-wave part of the amplitude, whereas in our case the
$s$-wave of the $D_{13}\to\pi\Delta$ vertex (see Table~\ref{tab2})
remains constant. We will return to this question when the low-energy
behavior of the total cross section will be discussed. 

For a qualitative analysis of the resonance contributions we 
write the corresponding total cross section in a simplified form 
\begin{equation}\label{37}
\sigma\approx\frac{\pi}{4\omega_\gamma^2}\sum_{N^*}(2J+1)
\Gamma_{\gamma N^*}(W)|G_{N^*}(W)|^2
\Gamma_{N^*\to\pi\pi N}(W)\,,
\end{equation}
where the total two-pion decay width $\Gamma_{N^*\to\pi\pi N}$ is
written as an incoherent sum of the contributions of the various
intermediate quasi-two-body channels, i.e.
\begin{equation}
\Gamma_{N^*\to\pi\pi N}=\Gamma_{N^*\to\pi\pi N}^{(\pi\Delta)}
+\Gamma_{N^*\to\pi\pi N}^{(\rho N)}+\Gamma_{N^*\to\pi\pi N}^{(\sigma N)}\,,
\end{equation}
which is correct provided one can neglect the overlap between the
different resonance amplitudes in~(\ref{25}). One can expect a
relatively small interference at least between $t^{\pi\Delta}$ and
$t^{\rho N}$ because of a relatively small width of the $\Delta$ and
$\rho$ resonances compared to the mass splitting between the 
different channels (in the limit of vanishingly small widths the final
quasi-two-body states should be orthogonal). Therefore, although for
an exact partial wave analysis the overlap between quasi-two-body
channels in the observables should be taken into account, this
interference is omitted for the moment being. Using Eq.~(\ref{37}),
the contribution of an individual resonance $N^*(J,L)$ at $W=M_{N^*}$
is estimated as 
\begin{equation}\label{42}
\sigma^{N^*}
\approx(2J+1)\,\frac{\pi}{\omega_\gamma^2}\,\frac{\Gamma_{\gamma
N^*}\Gamma_{N^*\to\pi\pi N}}{\Gamma_{tot}^2}\,.  
\end{equation}
It may be instructive to supplement Eq.~(\ref{42}) by the separate 
contributions of the intermediate quasi-two-body states
to the various charge channels of two pion production. Factoring out
and evaluating explicitly the isospin matrix elements, one obtains 
for $T=\frac32$ resonances  
\begin{eqnarray}\label{42a}
\sigma^{N^*}(\pi^+\pi^-)&=&\frac{26}{45}\sigma_{\pi\Delta}^{N^*}
+\frac{2}{3}\sigma_{\rho N}^{N^*}\,,\nonumber\\
\sigma^{N^*}(\pi^+\pi^0)&=&\frac{17}{45}\sigma_{\pi\Delta}^{N^*}
+\frac{1}{3}\sigma_{\rho N}^{N^*}=\sigma^{N^*}(\pi^-\pi^0)\,,\\
\sigma^{N^*}(\pi^0\pi^0)&=&\frac{2}{45}\sigma_{\pi\Delta}^{N^*}\nonumber\,,
\end{eqnarray}
and for $T=\frac12$ resonances
\begin{eqnarray}\label{42b}
\sigma^{N^*}(\pi^+\pi^-)&=&\frac{5}{9}\sigma_{\pi\Delta}^{N^*}
+\frac{1}{3}\sigma_{\rho N}^{N^*}+\frac{2}{3}\sigma_{\sigma N}^{N^*}
\,,\nonumber\\
\sigma^{N^*}(\pi^+\pi^0)&=&\frac{2}{9}\sigma_{\pi\Delta}^{N^*}
+\frac{2}{3}\sigma_{\rho N}^{N^*}=\sigma^{N^*}(\pi^-\pi^0)\,,\\
\sigma^{N^*}(\pi^0\pi^0)&=&\frac{2}{9}\sigma_{\pi\Delta}^{N^*}
+\frac{1}{3}\sigma_{\sigma N}^{N^*}\,. 
\nonumber
\end{eqnarray}
Here the partial cross sections $\sigma^{N^*}_X$ 
($X\in\{\pi\Delta,\rho N,\sigma N\}$) are obtained from Eq.~(\ref{42})
by substituting the total $\pi\pi N$ width $\Gamma_{N^*\to\pi\pi N}$
by the partial width $\Gamma^{(X)}_{N^*\to\pi\pi N}$. As discussed
above, these relations are exact only for vanishing overlap 
between the quasi-two-body channels, and in practice should be used
for qualitative estimates only. On the other hand, a comparison of the
straightforward evaluation shows that the approximations (\ref{42})
through (\ref{42b}) reproduce indeed quite well the actual resonance
contribution to the total cross section. Only for those resonances,
which are strongly coupled to $\sigma N$ channel the cross section is
sensitive to the sign of the $N^*\to\sigma N$ coupling, primarily because
of a large width of the $\sigma$ meson. Evaluating (\ref{42b}) for the Roper
resonance $P_{11}(1440)$, one finds a contribution of about 0.8~$\mu$b
to the $\pi^0\pi^0$ cross section at $E_\gamma=635$~MeV, which totally
excludes its dominance in this channel. This result is consistent with
the calculation of \cite{Oset} and is at variance with the theoretical
prediction of~\cite{GRAAL}. The role of the $S_{11}(1535)$ resonance,
which is often ignored in $\pi\pi$ photoproduction models, is almost
as important as that of $P_{11}(1440)$. Furthermore, comparing
(\ref{42a}) with (\ref{42b}) for the $\pi^0\pi^0$ channel, one readily
sees that the contribution of $T=3/2$ resonances to this channel
should be small in general. 

As is shown in Figs.~\ref{fig2} through \ref{fig2b}, 
the $D_{13}(1520)$ provides the dominant resonance contribution to
double pion photoproduction in the second resonance region. A
significant role of the $D_{33}(1700)$ and $F_{15}(1680)$ is also worth
noting. Other resonances are less pronounced. We would like to stress the
fact that we did not fit the resonance parameters to the observed cross
sections. Therefore, the quality of the description of the data in
Fig.~\ref{fig2} through \ref{fig2b} is not perfect. In particular we
do not reproduce the position of the second peak observed in the
$\pi^+\pi^-$ and $\pi^0\pi^0$ channels. Also the $\pi^+\pi^0$
experimental cross section is underestimated in the region below
$E_\gamma=0.7$~GeV. 

In addition, there is no room in our model for a large contribution of
the $D_{33}(1700)$ resonance. As was already noted in~\cite{Oset2},
having a large width and a strong coupling to the $s$-wave $\pi\Delta$
state, this resonance can interfere with the $\Delta$ Kroll-Ruderman
term and, therefore, influences the resulting cross section over a
wide energy region. One should notice that analogously to
$D_{13}(1520)$, the $D_{33}(1700)$ resonance tends to increase the
total cross section for $\pi^+\pi^-$ production, in contrast to the
results of~\cite{Oset2}, where its inclusion reduces the $\pi^+\pi^-$
cross section. The analysis of~\cite{Manley} gives opposite signs for
the decay amplitudes $D_{13}(1520)\to\pi\Delta$ and
$D_{33}(1770)\to\pi\Delta$ with respect to the corresponding $\pi N$
decay amplitudes. But from the multipole analysis of pion
photoproduction follows that the $\pi N$ vertices of these resonances
have also opposite signs (see Table~\ref{tab2}), so that the total
phase of both amplitudes is the same. As one can see in
Fig.~\ref{fig2}, including $D_{33}(1700)$, one obtains a sizeable
overestimation of the data. We cannot, however, conclude that our
evaluation tends to favor a less important role of this resonance in
double pion photoproduction than the one using the parameters given
in~\cite{PDG}. More likely, this inconsistency points to shortcomings
in the Born amplitude, whose role may well be overestimated by the
present model. 

For $\pi^0\pi^0$ photoproduction, our calculation predicts a second
peak at $E_\gamma=1$~GeV mostly coming from the excitation of the
$F_{15}(1680)$ resonance. Its position is in rough agreement with
experiment~\cite{GRAAL}, but its magnitude is considerably
underestimated. However, one can obtain a much better description of
the experimental cross section in this region (dotted curve in
Fig.~\ref{fig2b}) by choosing instead of a negative sign of the
$F_{15}\pi\Delta$ coupling from~\cite{Manley} a positive one as
predicted in~\cite{Capstic}. In Ref.~\cite{GRAAL}, the second peak in
the $\pi^0\pi^0$ data was explained as an interference effect between
the $P_{11}\to\sigma N\to\pi^0\pi^0 N$ mechanism and a very strong
background from photoproduction of a $\sigma$ meson via $\rho$
exchange (diagram (7) of Fig.~\ref{fig1}). However, in the present
model, the background from intermediate $\sigma$ photoproduction, for
which our model predicts about 0.5~$\mu$b at $E_\gamma=1$~GeV, turns
out to be rather insignificant. We did not include in the calculation
the $P_{11}$(1710) resonance, but a simple estimate, using the
expressions of (\ref{42}) and (\ref{42b}), gives for the
$P_{11}(1710)\to\sigma N\to\pi^0\pi^0 N$ mechanism only about
0.3~$\mu$b at the same energy, so that the combined contribution from
these terms does not strongly influence the $\pi^0\pi^0$ cross
section. 

In Fig.~\ref{fig2rho} we show the cross section for $\rho$-meson
photoproduction. In this case we have extended the
calculation up to $E_\gamma=3$~GeV in order to show the trend of the
cross section at higher energies. In $\rho^0$ meson production, the
charge conjugation invariance forbids the exchange of vector mesons,
so that only spin-zero mesons can contribute to $t$-channel
exchange. As is shown in Fig.~\ref{fig2rho}, within the present model,
most of neutral $\rho$ production comes from $\sigma$-exchange
(diagram (6) in Fig.~\ref{fig1}), except for the ``subthreshold''
region, i.e.\ $E_\gamma < 1$~GeV, which is dominated by baryon
resonance excitations. The role of $\pi$-exchange remains
insignificant, primarily, because of a very weak coupling at the
$\gamma\pi^0\rho^0$ vertex. Although the dominance of the
$\gamma(\pi\pi)$ mode in the radiative $\rho^0$-meson decay can serve
as a strong justification for the $\sigma$-exchange model, its status
in the theory of $\rho^0$ production is not clear. There are more
refined models (see e.g.~\cite{Oh} and references therein) which are
however much more complicated. 

Without the $D_{33}(1700)$ resonance, one obtains a satisfactory
agreement with the cross section data for $\pi^+\pi^-$ and
$\pi^+\pi^0$ production, but there is a sizeable deviation for the
$\pi^0\pi^0$ channel. In the region up to $E_\gamma < 0.7$~GeV, the
theoretical results ly far below the experimental points so that the
data can not be fitted simply by varying the parameters of
resonances. The relatively steep rise of the experimental cross section 
in Fig.~\ref{fig2b} right above the threshold indicates quite a 
strong $s$-wave contribution to the production amplitude, which, however, 
is not born out by our model. But in order to reach a more definite 
conclusion with respect to the partial wave structure of the
amplitude, one needs experimental angular distributions of the
produced particles. Of the various Born amplitudes, which contribute
to the $s$-wave part, only the $N\Delta$ crossed term (the second
diagram of (12) of Fig.~\ref{fig1}) appears to be relatively important
at low energies (dash-dotted curve in the left panel of
Fig.~\ref{fig2b}), but its contribution is, however, not sufficient to
explain the observed cross section. Also the so-called $Z$-graphs do
not seem to play a sizeable role. As an example, we demonstrate in
Fig.~\ref{fig2b} the importance of the diagram (17) where the final
$\pi^0\Delta^+$ state can be produced in an $s$-wave via
photoabsorption on an antinucleon. It is the only possibility to leave
the $\pi\Delta$ system in a relative $s$-state. Other possible
couplings with an antinucleon lead to higher partial $\pi\Delta$ waves
and should be much less important near threshold. As one can see from
Fig.~\ref{fig2b}, the $Z$-graph plays only a secondary role and cannot
explain the rapid rise of the experimental cross section just above 
threshold. 

With respect to the resonance mechanisms, our calculation predicts a
relatively small contribution from $D_{13}(1520)$ around a photon
energy of $E_\gamma=0.6$~GeV, which is about three times smaller 
than the one in Ref.~\cite{Oset}. The possible reason for this difference
was already partially explained above. Namely, the negative $q^2$-dependent
contribution to the $s$-wave amplitude in~\cite{Oset} requires on 
the other hand a much stronger momentum independent part at low
energy than in our model, thus making the energy dependence of the 
cross section less pronounced. Here we do not discuss such effects like 
$\pi^+\pi^-\to\pi^0\pi^0$ scattering, considered in~\cite{Oset1} at very 
low energies, as well as the influence of the type of $\pi NN$ coupling, 
discussed in~\cite{Hirata}. In any case, because of such a strong
model dependence of the low energy $\pi^0\pi^0$ cross section, more
refined theoretical and experimental investigations 
of this channel are needed. In
general, it should be noted, that there is no qualitative agreement
between different authors with respect to the contribution of the
$D_{13}(1520)$ resonance. Even at the resonance peak around
$E_\gamma=0.75$~GeV, where the value of the cross section is fixed
almost unambiguously by the electromagnetic and hadronic decay widths
(see Eq.~(\ref{42})), the size of the cross section varies very
strongly. For example, in the $\pi^+\pi^0$ channel, the contribution
of the $D_{13}(1520)$ is about 30~$\mu$b in~\cite{Ochi} but only
20~$\mu$b in~\cite{Oset} and in our work. 

As for the reaction on the neutron, one notes in Fig.~\ref{fig6} that
the $\pi^+\pi^-n$ and $\pi^-\pi^0n$ cross sections are practically
equal to the corresponding $\pi^+\pi^-p$ and $\pi^+\pi^0p$ cross
sections. Comparison with the old data for
$\pi^+\pi^-n$~\cite{Carbonara} shows that the theory gives values
systematically higher than the experimental results by about 20~\% at
the maximum of the cross section.

In Fig.~\ref{fig5} we present our results for the beam-target 
helicity asymmetry of the total cross section 
\begin{equation}\label{45}
\Delta\sigma=\sigma_{3/2}-\sigma_{1/2}\,,
\end{equation}
where $\sigma_\lambda$ corresponds to the total cross section for
parallel orientation of photon and target spins for
$\lambda=3/2$ and to the antiparallel orientation for $\lambda=1/2$. 
In comparison to our previous calculation~\cite{ArFS}, inclusion  of higher 
resonances as well as a more refined treatment of the Born terms shifts 
$\Delta\sigma$ above the $D_{13}(1529)$ peak to negative values, especially in 
the $\pi^+\pi^-$ and $\pi^0\pi^0$ channels. Again, in the low energy region 
we find a significant deviation of our results for $\pi^0\pi^0$ production 
from the preliminary data of~\cite{refmami}. In order to analyze 
the present results, we write the resonance contribution to the asymmetry
(\ref{45}) in the simplified form 
\begin{equation}\label{50}
\Delta\sigma\approx\frac{\pi}{2\omega_\gamma^2}\sum_{N^*}(2J+1)
\{\Gamma_{\gamma N^*}^{3/2}(W)  
-\Gamma_{\gamma N^*}^{1/2}(W)\}|G_{N^*}(W)|^2\Gamma_{N^*\to\pi\pi N}(W)\,,
\end{equation}
which is obtained under the same assumptions as (\ref{37}). The partial 
widths $\Gamma_{\gamma N^*}^{j}(W)$ ($j=1/2,\,3/2$) 
are defined in Appendix~\ref{AppB} in (\ref{A75}). 
The $D_{13}(1520)$ resonance, for which one finds
$\Gamma_{\gamma N^*}^{3/2}\approx 14\, \Gamma_{\gamma N^*}^{1/2}$~\cite{PDG} at
$W=M_{D_{13}}$ on the proton, contributes thus almost exclusively to
the $3/2$ part. However, as one can see from Fig.~\ref{fig5}, the
positive contribution of $D_{13}$ to $\Delta \sigma$ is more than 
canceled by a negative contribution of the $P_{11}$ resonance 
at low energies. Furthermore, the
Born terms alone exhibit a very small helicity dependence resulting in a
strong cancellation between $\sigma_{3/2}$ and $\sigma_{1/2}$ (dashed 
curve in the lower right panel of Fig.~\ref{fig5}). As a result, we
obtain an essentially negative value for the asymmetry (\ref{45})
around $E_\gamma=0.6$~GeV. This is in disagreement with the
experimental results obtained at MAMI~\cite{refmami} as well as with
the calculation of \cite{Hirata} where this asymmetry amounts at this
energy up to about +5~$\mu$b. Probably, the crucial origin of this
disagreement lies in the mentioned strong model dependence of the Born
sector. 

Finally, Fig.~\ref{fig5a} shows the result of our calculation of 
the beam asymmetry $\Sigma$ for linearly polarized photons in the 
$\pi^0\pi^0$ photoproduction, which is compared to the 
recent GRAAL data~\cite{GRAAL}. Although in the first energy bin one notes 
satisfactory agreement, one clearly sees for the higher energies an 
increased deviation from the experimental results which exhibit a 
slightly negative asymmetry in contrast to a positive asymmetry of the 
theory.

%% file: pap3_ch3.tex
%%%%%%%%%%%%%%%%%%%%%%%%%%%%%%%%%%%%%%%%%%%%%%%%%%%%%%%%%%%%%%%%%%%%%%%%%%%
\section{Double pion photoproduction on the deuteron}\label{sect3}

We will now turn to incoherent two-pion photoproduction on the deuteron 
\begin{equation}\label{55}
\gamma(k,\vec{\epsilon}_\lambda)
+d(p_d)\rightarrow \pi^{\mu_1}(q_1)+\pi^{\mu_1}(q_2)+N_1(p_1)+N_2(p_2)\,,
\end{equation}
The corresponding coherent process (without deuteron break up) in the
"neutral" channels $\pi^+\pi^-$ and $\pi^0\pi^0$ has a very small cross
section and will be considered very briefly at the end of the section. 

The reaction (\ref{55}) is in principle considerably more complex than
the reaction on the nucleon, because in addition to the production on
each of the two nucleons one would have to consider electromagnetic
two-body production operators. The latter, however, will be neglected
in the present work. Thus the e.m.\ interaction consists in the sum of
the one-body production operators, which often is called the impulse
approximation (IA) if in addition the interaction between the various
particles in the final state (FSI) is neglected. But in the present
work, we will at least include the interaction between the two final
nucleons. 

In the $\gamma d$ c.m.\ system, the spin-averaged cross section of 
the reaction (\ref{55}) then reads 
\begin{equation}\label{60}
d\sigma=(2\pi)^{-8}\frac{E_dM_N^2Qp^*q^*}{8W^2\omega_\gamma}\ 
\frac{1}{6}
\sum\limits_{\lambda,M_d}\sum\limits_{S,M_S}\bigg|\sum\limits_{I=0,1} 
T_{IM_I SM_S\lambda M_d}^{\mu_1\mu_2}
(\vec q_1,\vec q_2,\vec{p},\vec k\,)\bigg|^2\,
d\Omega_Qd\omega_{NN}d\omega_{\pi\pi}d\Omega_{p^*}d\Omega_{q^*}\,, 
\end{equation}
where $\vec{Q}=\vec{q}_1+\vec{q}_2$ denotes the total 3-momentum of the 
two final pions, $\omega_{\pi\pi}$ the invariant $\pi\pi$ mass, $\vec
q^{\,\ast}$ the relative momentum in the $\pi\pi$ restsystem, and
$\omega_{NN}$ and $\vec p^{\,\ast}$ the corresponding quantities for the 
$NN$ subsystem, whereas $\vec p$ denotes the relative two-nucleon momentum
in the overall $\gamma d$ c.m.\ system. 
Furthermore, $T$ stands for the reaction matrix, in 
which the e.m.\ interaction is represented by the the sum of
the two-pion production amplitudes on each of the nucleons, i.e.
\begin{equation}
T_{IM_I SM_S\lambda M_d}^{\mu_1\mu_2}(\vec q_1,\vec q_2,\vec{p},\vec k\,)=
{^{(-)}\langle} \vec{p}\,;
SM_S; IM_I|\Big(\hat t^{\mu_1\mu_2,(1)}_\lambda(\vec q_1,\vec q_2,\vec k\,)
+\hat t^{\mu_1\mu_2,(2)}_\lambda(\vec q_1,\vec q_2,\vec k\,)\Big)|1M_d;00
\rangle\,,
\end{equation}
where $t^{\mu_1\mu_2,(i)}_\lambda$ denotes the elementary production operator 
on nucleon ``$i$''.

The final two-nucleon state, characterized by the total spin $S$ and
isospin $I$ and the corresponding projections $M_S$ and $M_I$, 
is represented by a
relative scattering wave $|\vec{p}\,;SM_S;IM_I\rangle^{(-)}$ 
which is determined by the $NN$ scattering matrix $t_{NN}$ 
\begin{equation}\label{NNscattering}
|\vec{p}\,;SM_S;IM_I\rangle^{(-)}=
\Big(1+t_{NN}\,G_{NN}(W_{NN}-i\epsilon)\Big)\frac{1}{\sqrt{2}} 
\Big[|\vec{p}\,\rangle-(-1)^{S+I}|-\vec{p}\,\rangle\Big]
|SM_S\rangle|IM_I\rangle\,,
\end{equation}
where $|\vec{p}\,\rangle$ denotes a relative 
two-nucleon plane wave and $G_{NN}$ the
free relative $NN$ propagator.

According to the two contributions to the final $NN$ state in
(\ref{NNscattering}) the reaction matrix $T$ may be split into two 
terms, the standard impulse approximation (IA) or spectator model as
the basic part and the rescattering contribution of the
nucleons in the final state, so that the resulting amplitude is given by
\begin{equation}\label{70}
T_{IM_I SM_S\lambda M_d}^{\mu_1\mu_2}(\vec q_1,\vec q_2,\vec{p},\vec k\,)=
T_{IM_I SM_S\lambda M_d}^{\mu_1\mu_2,\,(IA)}
(\vec q_1,\vec q_2,\vec{p},\vec k\,)
+T_{IM_I SM_S\lambda M_d}^{\mu_1\mu_2,\,(NN)}
(\vec q_1,\vec q_2,\vec{p},\vec k\,)\,.
\end{equation}
Denoting the matrix element of the elementary production on a nucleon with 
initial momentum $\vec p_i$ by
$t_\lambda^{\mu_1\mu_2}(\vec q_1,\vec q_2,\vec p_i,\vec k\,)$, 
the amplitude of the IA-term reads
\begin{eqnarray}\label{75}
T_{IM_I SM_S\lambda M_d}^{\mu_1\mu_2,\,(IA)}
(\vec q_1,\vec q_2,\vec{p},\vec k\,)
&=& \sqrt{2}\sum\limits_{M_S'}\langle IM_I|\langle SM_S|\Big[
t_\lambda^{\mu_1\mu_2}(\vec q_1,\vec q_2,\frac12\vec Q+\vec{p}-\vec k,\vec k\,)
%\langle\vec{p}-\frac12 \vec Q\,|{t}^{(1)}_\lambda
%|\vec{p}+\frac12 \vec Q-\vec k\rangle
\Psi_{M_S'M_d}\Big(\frac{1}{2}(\vec{Q}-\vec{k})+\vec p\,\Big)
\nonumber\\
&& %\hspace*{-1.5cm}
-(-1)^{S+I}
t_\lambda^{\mu_1\mu_2}(\vec q_1,\vec q_2,\frac12\vec Q-\vec{p}-\vec k,\vec k\,)
%\langle-\vec{p}+\frac12 \vec Q\,|{t}^{(1)}_\lambda
%|-\vec{p}+\frac12 \vec Q-\vec k\,\rangle
\Psi_{M_S'M_d}\Big(\frac{1}{2}(\vec{Q}-\vec{k})-\vec p\,\Big)
\Big]|1M_S'\rangle|00\rangle\,,
\end{eqnarray} 
where $\Psi_{M_SM_d}$ represents the component of the deuteron wave
function with a definite two-nucleon spin projection $M_S$
\begin{equation}\label{73}
\Psi_{M_SM_d}(\vec{p}\,)=\sum\limits_{L=0,2}(LM_L\,1M_S|1M_d)
u_L(p)Y_{LM_L}(\hat{p})\,.
\end{equation}
The rescattering term in (\ref{70}) is given by
\begin{eqnarray}\label{80}
T_{IM_I SM_S\lambda M_d}^{\mu_1\mu_2,\,(NN)}(\vec q_1,\vec q_2,\vec{p},\vec k\,)
&=&\sum\limits_{M_S'}\int\frac{d^3p'}
{(2\pi)^3}\,
t_{IM_I S,M_S'M_S}^{NN}(\vec p,\vec{p}^{\,\prime}\,)\,
G_{NN}(W_{NN})\, %\nonumber \\&& 
T_{IM_I SM_S\lambda M_d}^{\mu_1\mu_2,\,(IA)}
(\vec q_1,\vec q_2,\vec{p}^{\,\prime},\vec k\,)\,,
\end{eqnarray} 
where $t^{NN}$ denotes the half off-shell $NN$ scattering matrix.
In the calculation of the integral in (\ref{80}) we use a partial wave
decomposition of the $NN$ scattering states.  

The isospin structure of the amplitude is easily evaluated using 
(\ref{20}) with the isospin operators listed in (\ref{A5}) through 
(\ref{iso_O3pm})
\begin{eqnarray}\label{85}
\langle 00|t^{\mu_1\mu_2}|00\rangle&=&\delta_{\mu_1,-\mu_2}
\Big((-1)^{\mu_1}f_1^{(+)}+\mu_1 f_2^{(-)}\Big)\,,\\
\langle 1M_I|{t}^{\mu_1\mu_2}|00\rangle&=&
\delta_{M_I,0}\delta_{\mu_1,-\mu_2}
\Big((-1)^{\mu_1}f_2^{(+)}-\mu_1f_1^{(-)}\Big)
+\frac12\Big((-1)^{\mu_2}\delta_{\mu_1,0}\delta_{M_I,-\mu_2}
+(\mu_1\leftrightarrow\mu_2)\Big)f_3^{(+)}\nonumber\\
&&-\Big(\mu_1\delta_{\mu_2,0}\delta_{M_I,-\mu_1}
-(\mu_1\leftrightarrow\mu_2)\Big)f_1^{(-)} 
-\frac12\Big((-1)^{\mu_2}\delta_{\mu_1,0}\delta_{M_I,-\mu_2}
-(\mu_1\leftrightarrow\mu_2)\Big)f_3^{(-)}\,.\label{85b}
\end{eqnarray}

Since calculational details associated with the evaluation of the 
two-nucleon interaction in incoherent meson production on the deuteron were 
considered in many papers (see, e.g.~\cite{Laget2,Levchuk,Darw}) 
there is no need to repeat them here. We only note that $S$, $P$, 
and $D$ waves were included in the $NN$ scattering matrix, for which 
the separable version of the Paris potential from \cite{Paris} was 
used. For the sake of consistency, also the deuteron wave function 
was calculated using the separable form of the potential. 

As our main result, we present in Fig.~\ref{fig7} the total cross
sections for the various charge configurations of 
$\pi\pi$ photoproduction on the deuteron. 
The cross sections reproduce qualitatively the form
of the elementary cross section (dotted curves) except that they are
slightly smeared out by the Fermi motion especially around the resonance 
peaks. This fact together with the obviously quite small influence 
of $NN$ rescattering in the final state support the approximate validity of 
the spectator model. Even in the "neutral" $\pi^+\pi^-$ channel, which is 
by far the most important channel, FSI leads to a lowering by
only 10~\% of the plane wave cross section. Thus it is significantly smaller 
than what had been found in single neutral pion production on the deuteron
$\gamma d\to\pi^0 np$ where this effect lead to a reduction 
by about 30~\%. 

This feature is easily explained by
the relatively large momentum transfer associated with the production of two
pions. Firstly, it leads to a reduction of the region of
small distances between the nucleons, where the $NN$
interaction is sizeable. Furthermore, more importantly in the neutral
channel where also the coherent transition (without deuteron break up) is
possible, is the nonorthogonality of the initial and
final $NN$ wave functions in IA. As a consequence, the IA contains 
part of the coherent reaction. The size  
of this "nonorthogonal contribution" is roughly given by the  
coherent cross section (see e.g.~\cite{Kolyb}) and depends 
strongly on the momentum transfer to the $NN$ subsystem 
(in the extreme case when the momentum
transfer goes to zero, the IA contains it completely). 
In particular, this effect leads to a large 
role of $NN$ FSI in the single $\pi^0$ photoproduction on the deuteron where 
the coherent 
channel turns out to be quite sizeable. Again, the role of orthogonality in
the $\pi\pi$ reactions is reduced because of a significantly increased
momentum transfer. 
Comparison with the available data in Fig.~\ref{fig7} shows that 
the agreement in the $\pi^+\pi^-$ and $\pi^-\pi^0$ channels is quite
satisfactory. Deviation from the $\pi^0\pi^0$ data should arise from 
the same origin as that discussed above for the corresponding 
elementary reaction.

The coherent cross sections in the 
$\pi^+\pi^-$ and $\pi^0\pi^0$ channels are presented 
in Fig.~\ref{fig8}. In contrast to the 
single pion case, the coherent $\pi^+\pi^-$ photoproduction 
comprise only about 6~\% of the corresponding 
incoherent cross sections in Fig.~\ref{fig7}. 
The $\gamma d\to\pi^0\pi^0 d$ cross section turns out to be vanishingly small. 
In the last 
case only the symmetric part, proportional to $f_1^{(+)}$ in (\ref{85})
contributes. For the resonances with isospin $T$=1/2 it is determined by the 
isoscalar part of the $\gamma N\to N^*$ transition (see Table \ref{tab4}), 
which is small for almost all resonances considered here. For $T$=3/2 
resonances as well as for most of the Born terms, which are important in the 
elementary $\pi^0\pi^0$ photoproduction, the amplitude $f_1^{(+)}$ 
is zero (see Table \ref{tab4}). As a result, the coherent $\pi^0\pi^0$ cross 
section comprises about 0.5 $\%$ of the incoherent one.

Finally. we present in Fig.~\ref{fig7a} the beam-target helicity 
asymmetry $\Delta\sigma$, defined analogously as the one for the 
elementary reaction on a nucleon (\ref{45}),
\begin{equation}
\Delta\sigma=\sigma^P-\sigma^A\,,
\end{equation}
where $\sigma^{P/A}$ correspond to the total cross section for
parallel/antiparallel orientation of photon and target spins, respectively. 
As in the elementary case, these asymmetries exhibit positive maxima
around 700-750~MeV corresponding to the contribution of the $D_{13}(1520)$
resonance and then decrease rapidly with increasing energy towards negative
values above roughly 1.2~GeV. Compared to our earlier evaluation~\cite{ArFS}, 
this feature reduces substantially the two-pion contribution to the 
Gerasimov-Drell-Hearn (GDH) integral. The results of the explicit 
integration up to 1.5~GeV are listed in Table~\ref{tabGDH} together
with the finite total GDH-integral including $\pi$, $\eta$, and $\pi\pi$ 
contributions, and photodisintegration in case of the deuteron. For the neutron
one notices a substantial underestimation of the GDH value by about 40~$\mu$b, 
while for the proton only a slight overestimation is found. For the deuteron
a positive contribution of about 29~$\mu$b is missing. 

Concluding this section we would like to comment on some restrictions of our
results. As is stressed above, the corrections from the $NN$ interaction are
relatively small. However, the conclusion about the general validity
of the spectator model for the photoproduction on the deuteron has to be
confronted with the experimental results of \cite{Asai}. In this paper the
transition $\gamma d\to\Delta^{++}\Delta^-$ with subsequent decay 
to the $pn\pi^+\pi^-$ state was extracted, using the analysis of the $\pi N$
invariant mass distributions. The crucial point is that for this transition 
to happen both nucleons have to participate actively in the reaction, 
quite in contrast to the
present treatment, where the second nucleon takes part only in the distortion
of the final $NN$ waves. According to the measured yields, the two-body 
mechanism
provides about 30~\% of the total cross section, thus making the validity of
the spectator model very doubtful. This conclusion was confirmed by the
calculation presented in~\cite{Oset3} where the transition to the 
$\Delta^{++}\Delta^-$ state amounts to 40~$\mu$b at $E_\gamma=800$~MeV.  
Assuming a two-step mechanism one could naively estimate that it would lead to 
a contribution comparable in size with pion rescattering in single 
pion photoproduction on the deuteron. However, this last effect was 
shown to be vanishingly small~\cite{Laget2,Levchuk}. Thus we think, 
this aspect deserves a more detailed study. 

%% file: pap3_ch4.tex
\section{Summary and conclusions}

In the present paper, we have extended the elementary two-pion 
photoproduction operator used in 
previous work~\cite{Oset,Laget,Ochi} to higher energies,  
including all four-star resonances with masses $M < 1.8$ GeV.
This operator is based on an effective Lagrangean approach evaluating only
tree-level diagrams. The necessary coupling strengths are determined 
by the hadronic and electromagnetic decays of the resonances. 
The present approach does not allow a high precision description of 
the reaction, primarily because of a nonrelativistic treatment of
the baryons and other shortcomings. However, it should be able to 
account for the main features of the reaction so that at least 
qualitative conclusions about the underlying mechanisms can be drawn. 

In the $\pi^+\pi^-$ and $\pi^+\pi^0$ channels  
it is quite easy to obtain a decent description of the data, 
but only if the large contribution of $D_{33}(1700)$ is excluded. 
Furthermore, we face principal difficulties in the $\pi^0\pi^0$ channel, where 
the theory strongly underestimates the cross section data in the near 
threshold region below $E_\gamma$=0.7 GeV. It should be stressed that 
from all charge channels the $\pi^0\pi^0$ one is the least understood, 
but it is also the weakest one, less than 15~\% of the dominant 
$\pi^+\pi^-$ channel. Apart from the noted underestimation at low 
energies, there is also a visible deviation between experimental 
and theoretical results for the $\Sigma$ 
asymmetry for linearly polarized photons 
as well as for the $\pi^0\pi^0$ photoproduction on a neutron as was 
noted in \cite{GRAAL} and \cite{Kleber}. 
We think that more detailed studies of angular distributions as
well as polarization measurements can help to clarify the role of different
mechanisms in the $\pi\pi$ photoproduction in the second resonance region. 
In particular, the investigation of the beam asymmetry
with the CLASS detector~\cite{Strauch} seems to be very promising for studying 
the structure of the production amplitude.

A major unresolved problem is the role of final state interaction in the
$\pi\pi N$ system which is also closely connected to the problem of 
unitarity and analyticity of the overall production amplitude. 
In the present paper we took into account only quite
roughly the effect of absorption in the final state within the WKB
prescription. Although this quasi-classical method is used also at low
energies, its validity should be doubted in the second resonance region, 
where only the lowest partial waves dominate the cross section. 
As an alternative approach, the method used in~\cite{Mokeev} treats 
the $\pi NN$ interaction as an effective
quasi-two-body scattering via resonance excitations. But we think, it is not
clear whether such a simplified approach to the $\pi\pi N$ system can account
for its quite complicated dynamics. For a more realistic
treatment of the $\pi\pi N$ interaction, a rigorous three-body scattering
approach is needed, where not only two-body but also three-body unitarity 
of the $\pi\pi N$ interaction can
be systematically incorporated. Because of quite a large amount of resonances
involved in $\pi N$ scattering, a full analysis of the $\pi\pi N$ system 
appears to be quite complex. However, a first and apparently 
very good approximation is provided by taking into account only 
the $\Delta$(1232) resonance in the $\pi N$ and and the $\sigma$-resonance 
in the $\pi\pi$ subsystem. 

As for the reaction on the deuteron, 
our primary aim was to investigate the role of the $NN$ 
interaction between the final nucleons, thus testing the validity of the
spectator model. The 
main result is that the effect of the $NN$ interaction is quite small. 
From this point of view, the spectator model can be considered as a good first 
approximation for a rough determination of the
elementary neutron amplitude from the deuteron data in the region of quasifree
kinematics. However, for precision studies, although at present out of sight,
such FSI and other nuclear effects have to be considered with care. 

Concluding the paper, we would like to mention, that in our opinion, 
future work should be oriented not only to a refinement of the 
$\pi\pi$-photoproduction model, but more importantly to a unification 
of the models for single and double pion photoproduction. In other words, 
a consistent treatment providing at the same time an at least reasonable 
description for both channels would be more useful, than an excellent 
but independent fit of these two types of reactions, achieved with very 
different sets of parameter values.

%% file: pap3_app.tex
\appendix
\section{The isotopic spin structure of the $\gamma N\to \pi\pi N$ amplitude}
\label{AppA}
\renewcommand{\theequation}{A.\arabic{equation}}
\setcounter{equation}{0}

In this appendix we discuss the isotopic spin structure
of the amplitude for double pion photoproduction.
Our goal is to present the production amplitude as an operator in the
nucleon isotopic spin space, analogously to what is done for single pion
photoproduction~\cite{CGLN}, where the existence of three independent
charge channels is described by three corresponding isospin operators 
\begin{equation}\label{A0}
t(\gamma N\to \pi^\mu N)= A^{(0)}\tau^\dagger_{\mu}
+A^{(-)}\frac{1}{2}[\tau^\dagger_{\mu},\tau_0]
+A^{(+)}\frac{1}{2}\{\tau^\dagger_{\mu},\tau_0\}\,. 
\end{equation}
Our convention for the isospin operators are 
\begin{equation}
\tau_0=\tau_z\quad \mbox{and}\quad  
\tau_{\pm 1}=\mp\frac{1}{\sqrt{2}}(\tau_x\pm i\tau_y)\,,
\end{equation}
where $\vec \tau$ denotes the Pauli isospin operator, so that
\begin{equation}
\langle p|\tau_{1}|n\rangle=-\langle n|\tau_{-1}|p\rangle
=-\sqrt{2}\,,\quad
\langle p|\tau_0|p\rangle=-\langle n|\tau_0|n\rangle=1\,.
\end{equation}
 
In double pion production one has six independent charge channels,
which are described by six isospin operators ${\cal O}_{\mu_1\mu_2}$,
depending on $\tau_{\mu_1}^\dagger$ and $\tau_{\mu_2}^\dagger$.
Since the amplitude $t^{\mu_1\mu_2}$ has to be symmetric under
interchange of the two pions because of their boson property, i.e.\
\begin{equation}
t^{\mu_1\mu_2}(\vec q_1,\vec q_2,\vec k\,)= 
t^{\mu_2\mu_1}(\vec q_2,\vec q_1,\vec k\,)\,,
\end{equation}
the operators ${\cal O}_{\mu_1\mu_2}$ can be chosen with definite
symmetry property. It turns out that three of them are symmetric under
interchange $\mu_1\leftrightarrow\mu_2$, denoted by a superscript
``(+)'', and the other three antisymmetric with superscript ``($-$)''
\begin{eqnarray}\label{A5}
{\cal O}_{\mu_1\mu_2}^{1(+)}&=&
\frac{1}{2}\{\tau_{\mu_2}^\dagger,\tau_{\mu_1}^\dagger\}
=(-1)^{\mu_1}\delta_{\mu_1,-\mu_2}\,, \label{iso_O1p}\\
{\cal O}_{\mu_1\mu_2}^{1(-)}&=&
\frac{1}{2}[\tau_{\mu_2}^\dagger,\tau_{\mu_1}^\dagger]=
\sqrt{2}\,(1\mu_1\,1\mu_2|1\mu_1+\mu_2)\,\tau_{\mu_1+\mu_2}^\dagger\,,\\
{\cal O}_{\mu_1\mu_2}^{2(+)}&=&
\frac{1}{2}\{{\cal O}_{\mu_1\mu_2}^{1(+)},\tau_0\}
=(-1)^{\mu_1}\delta_{\mu_1,-\mu_2}\tau_{0}\,,\\
{\cal O}_{\mu_1\mu_2}^{2(-)}&=&
\frac{1}{2}\{{\cal O}_{\mu_1\mu_2}^{1(-)},\tau_0\}
=\mu_1\delta_{\mu_1,-\mu_2}\,,\\
{\cal O}_{\mu_1\mu_2}^{3(\pm)}&=&
\frac{1}{4}(\{\tau_{\mu_1}^\dagger,{\cal O}_{\mu_2\,0}^{2(+)}\}
\pm(\mu_1\leftrightarrow\mu_2))
=\frac12\left(
\delta_{\mu_2,0}\tau_{\mu_1}^\dagger\pm(\mu_1\leftrightarrow\mu_2)\right)\,.
\label{iso_O3pm}
\end{eqnarray}
Accordingly, the isotopic spin dependence of the amplitude is
represented by these operators with appropriate spin-spatial
amplitudes $f_l^{(\pm)}$, where the superscript denotes the symmetry
property under interchange of the two pions,
\begin{eqnarray}
t^{\mu_1\mu_2}(\vec q_1,\vec q_2,\vec k\,)
&=&\sum_{l=1}^3 (f_l^{(+)}(\vec q_1,\vec q_2,\vec k\,)
{\cal O}_{\mu_1\mu_2}^{l(+)}
+f_l^{(-)}(\vec q_1,\vec q_2,\vec k\,){\cal O}_{\mu_1\mu_2}^{l(-)})
\,.\label{isospinstructure}
\end{eqnarray}
The functions $f_l^{(\pm)}$ possess the symmetry property under the exchange
of the pion momenta
\begin{equation}
f_l^{(\pm)}(\vec{q}_1,\vec{q}_2,\vec k\,)=
\pm f_l^{(\pm)}(\vec{q}_2,\vec{q}_1,\vec k\,)\,.
\end{equation}
The symmetric amplitudes $f_l^{(+)}$ contribute to 
the isospin $T=0,2$ part of the $\pi\pi$ wave function and the antisymmetric 
ones $f_l^{(-)}$ to $T=1$.
For each diagram in Fig.~\ref{fig1} the dependence on the index $l$, 
i.e. on the operator type, is described by a coefficient so that 
the functions $f_l^{(\pm)}$ can be 
expressed in the simpler form 
\begin{equation}\label{A18}
f_l^{(\pm)}(\vec{q}_1,\vec{q}_2,\vec k\,)
=C_l^{(\pm)}f^{(\pm)}(\vec{q}_1,\vec{q}_2,\vec k\,)\,,
\end{equation}
where the coefficients $C_l^{(\pm)}$ are determined by the isotopic structure 
of the diagram. 
They are listed in Table~\ref{tab2} for the resonance terms as well as 
for several important Born terms.
The amplitudes $t^{\rho N}$ and $t^{\sigma N}$ as well as those Born 
terms, where the pions are produced
via the decay of intermediate mesons, contribute
either to antisymmetric or symmetric spin-spatial amplitudes only.
Thus one has
\begin{equation}\label{A20}
f^{(+/-)}(\vec{q}_1,\vec{q}_2,\vec k\,)=0\,, 
\end{equation}
for an intermediate $\rho/\sigma$ meson, respectively.
The remaining terms in Fig.~\ref{fig1}, contributing, e.g., to $t^{\pi\Delta}$ 
in (\ref{25}), have as a rule mixed spatial symmetry. 

For completeness we present also explicit expressions for $t^{\mu_1\mu_2}$ in 
terms of the functions $f_l^{(+)}$ for the different charge channels, 
which can be obtained from (\ref{A5}) through (\ref{isospinstructure})
\begin{eqnarray}
&&\langle p|t^{1-1}|p\rangle =
-f_1^{(+)}-f_2^{(+)}-f_1^{(-)}+f_2^{(-)}\,,\ \quad
\langle n|t^{1-1}|n\rangle = 
-f_1^{(+)}+f_2^{(+)}+f_1^{(-)}+f_2^{(-)}\,,\nonumber \\
&&\langle n|t^{10}|p\rangle\ =
-\frac{1}{\sqrt{2}}\left(f_3^{(+)}+2f_1^{(-)}+f_3^{(-)}\right)\,,\quad
\langle p|t^{-10}|n\rangle = 
\frac{1}{\sqrt{2}}\left(f_3^{(+)}-2f_1^{(-)}+f_3^{(-)}\right)\,,\\
&&\langle p|t^{00}|p\rangle\ =
\frac{1}{\sqrt{2}}\left(f_1^{(+)}+f_2^{(+)}+f_3^{(+)}\right)\,,\quad \quad \
\langle n|t^{00}|n\rangle = 
\frac{1}{\sqrt{2}}\left(f_1^{(+)}-f_2^{(+)}-f_3^{(+)}\right)\,.\nonumber
\end{eqnarray}
The expressions for $\pi^0\pi^0$ photoproduction contain an additional factor 
$1/\sqrt{2}$.

The inverse relations, expressing the amplitudes $f_l^{(\pm)}$ in terms of the 
various charge channels, read
\begin{eqnarray}\label{A25}
f_1^{(+)}&=&\frac{1}{\sqrt{2}}(t_p^{00}+t_n^{00})\,,\\
f_2^{(+)}&=&\frac{1}{2\sqrt{2}}(t^{10}+t^{-10})
-\frac{1}{2}(t_p^{1-1}-t_n^{1-1})\,,\\
f_3^{(+)}&=&\frac{1}{\sqrt{2}}(t_p^{00}-t_n^{00})
-\frac{1}{2\sqrt{2}}(t^{10}+t^{-10})
+\frac{1}{2}(t_p^{1-1}-t_n^{1-1})\,,\\
f_1^{(-)}&=&-\frac{1}{2\sqrt{2}}(t^{10}+t^{-10})\,,\\
f_2^{(-)}&=&\frac{1}{\sqrt{2}}(t_p^{00}+t_n^{00})
+\frac12(t_p^{1-1}+t_n^{1-1})\,,\\
f_3^{(-)}&=&\frac{1}{\sqrt{2}}(t^{-10}-t^{10})-f_3^{(+)}\,,
\end{eqnarray}
where we have introduced the notation
\begin{eqnarray}
t_{p/n}^{00}&:=& \langle p/n|t^{00}|p/n\rangle\,,\\
t_{p/n}^{1-1}&:=& \langle p/n|t^{1-1}|p/n\rangle\,,\\
t^{\pm10}&:=& \langle n/p|t^{\pm10}|p/n\rangle\,.
\end{eqnarray}

%%%%%%%%%%%%%%%%%%%%%%%%%%%%%%%%%%%%%%%%%%%%%%%%%%%%%%%%%%%%%%%%%%%%%%%%%
\section{The vertex functions}
\label{AppB}
\renewcommand{\theequation}{B.\arabic{equation}}
\setcounter{equation}{0}
Here we give detailed expressions for the isobar vertex
functions which determine the various diagrams in Fig.~\ref{fig1}. For
example, the transition $\gamma N\to N^*(\alpha)\to\pi\Delta\to\pi NN$
will have the form 
\begin{equation}\label{A30}
T=F_{\Delta\to\pi N}\,F_{N^*(\alpha)\to\pi\Delta}\,G_{\Delta}\,
G_{N^*(\alpha)}\,F_{\gamma N\to N^*(\alpha)}\,,
\end{equation}
where $F_{\gamma N\to N^*(\alpha)}$, $F_{N^*(\alpha)\to\pi\Delta}$ and
$F_{\Delta\to\pi N}$ denote the appropriate vertex functions for the
indicated transitions and $G_{N^*(\alpha)}$ and $G_{\Delta}$ the
corresponding intermediate dressed resonance propagators. The symbol
$\alpha$ stands for the resonance quantum numbers. In the following, we
will characterize a baryon resonance by its total angular momentum $J$
and its total $\pi N$ orbital momentum $L$ instead of its parity, 
i.e.\ $(\alpha)=(J,L)$. 

Before presenting the detailed formalism, we would like to clarify the 
notation. The vertex functions $F$ are given in terms of products of
irreducible tensors. For the irreducible tensor product we take the usual
definition of the coupling of two irreducible tensors
\begin{equation}\label{A35}
[Q^{[j']}\otimes P^{[j]}]^{[J]}_M=\sum\limits_{m'm}(j'm'\,jm|JM)
\,Q^{[j']}_{m'}P^{[j]}_m\,,
\end{equation} 
where $(j'm'\,jm|JM)$ denotes a Clebsch-Gordan coefficient. In
particular, the scalar product is related to the tensor product by
\begin{equation}\label{A40}
\big(Q^{[j]}\cdot P^{[j]}\big)= (-1)^{j}\sqrt{2j+1}\,
[Q^{[j]}\otimes P^{[j]}]^{[0]}=\sum\limits_m(-1)^{m}
\,Q^{[j]}_{m}P^{[j]}_{-m}\,.
\end{equation}
For $j=1$, Eq.~(\ref{A40}) defines the conventional scalar product of 
two three-vectors given in spherical coordinates, which are defined
for a vector $\vec q$ by
\begin{equation}\label{A45}
q_m=\sqrt{\frac{4\pi}{3}}{q}\,Y_{1m}(\hat{q})\,, \quad m=0,\pm 1\,.
\end{equation}
For a multiple product, i.e.\ a repeated coupling of a three-vector
$\vec Q$ with itself to the highest possible rank $l$, one 
obtains from (\ref{A40}) 
\begin{equation}\label{A50}
Q^{[l]}_m\equiv
[\dots[[Q^{[1]}\otimes Q^{[1]}]^{[2]}\otimes Q^{[1]}]^{[3]}\dots]^{[l]}
_m=\left(\frac{4\pi l!}{(2l+1)!!}\right)^{1/2}\,Q^lY_{lm}(\hat{Q})
\end{equation} 
Such multiple products of the type (\ref{A50}) appear in the resonance
couplings considered below. 

Furthermore, for the description of transitions 
$N(j=1/2)\to N^*(j')/\Delta(j')$ and 
$\Delta(j=3/2)\to N^*(j')$ one needs in general spin
transition operators $\sigma_{j'j}^{[J]}$. They are normalized so that
the corresponding matrix element is simply given by the associated
Clebsch-Gordan coefficient, i.e.\ 
\begin{equation}\label{A55}
\langle j'm'|\sigma_{j'j,\,M}^{[J]}|jm\rangle=(jm\,JM|j'm')\,
\quad \mbox{for}\quad j'\geq j\,.
\end{equation}
The spin operators for the inverse transitions are then determined 
by the conjugate operators 
\begin{equation}\label{A55a}
\sigma_{jj',M}^{[J]}=\sigma_{j'j,M}^{[J]\dagger}=(-1)^M\sigma_{j'j,-M}^{[J]}\,.
\end{equation}
It should be noted, that the operator $\sigma_{\frac12\frac12}$ as
defined above is not the ordinary Pauli spin operator but differs from
it by a factor $1/\sqrt{3}$, namely one finds 
\begin{equation}\label{A60}
\langle \frac12 m'|\sigma_{\frac12\frac12,\,M}^{[1]}|\frac12 m\rangle
=(\frac12m\,1M|\frac12 m')=\frac{1}{\sqrt{3}}\sigma_M\,,
\end{equation}
where $\sigma_M$ is a spherical component of the Pauli spin matrix.

Now we will consider the e.m.\ vertex functions possessing
electric and magnetic couplings. They can be presented in the general
form 
\begin{eqnarray}\label{A65a}
F_{\gamma N\to N^*(J,L)}^{E}&=&\frac{g^E}{2M_N^{j-1}}
\big(\sigma_{J,\frac12}^{[j]}\cdot[k^{[j-1]}\otimes\epsilon^{[1]}]^{[j]}
\big)\,, \quad j=2J-L\,,\\
F_{\gamma N\to N^*(J,L)}^{M}&=&\frac{g^M}{2M_N^j}
\big(\sigma_{J,\frac12}^{[j]}\cdot[k^{[j]}\otimes\epsilon^{[1]}]^{[j]}
\big)\,, \quad j=L \,,\label{A65b}
\end{eqnarray}
where $\vec k$ denotes the photon momentum, and the rank of the
multipole $j$ is fixed by the parity of the transition. For 
each resonance the values of $j$ are given in the last column 
of Table~\ref{tab1}. In order to fix the 
constants $g^E$ and $g^M$, we use the helicity 
amplitudes listed in~\cite{PDG}. They are related to the vertices in 
(\ref{A65a}) and (\ref{A65b}) by
\begin{equation}\label{A70}
A_\lambda(J,L)
=\frac{1}{\sqrt{2\omega_\gamma}}\langle N^*;J,\lambda)|
F_{\gamma N\to N^*(J,L)}^{E}+F_{\gamma N\to N^*(J,L)}^{M}
|N;\frac12,\lambda-1;\  
\gamma;(\lambda_\gamma=1)\rangle\,,\quad \lambda=\frac12,\,\frac32\,.
\end{equation}
The electromagnetic width of the decay of a resonance $N^*(J,L)$ into
a nucleon then is 
\begin{equation}\label{A75}
\Gamma_{N^*(J,L)\to\gamma N}=\frac{2M_N\omega_\gamma^2}{\pi(2J+1)M_{N^*}}
\left(|A_{1/2}(J,L)|^2+|A_{3/2}(J,L)|^2\right)=
\Gamma_{\gamma N^*}^{1/2}(W)+\Gamma_{\gamma N^*}^{3/2}(W)\,.
\end{equation}

From Eq.~(\ref{A70}) one obtains the following expressions relating
the constants $g^{E(M)}$ to the helicity amplitudes $A_\lambda$ 
\begin{equation}\label{A80}
A_\lambda(J,L)=\frac{1}{4\sqrt{\omega_\gamma}}
\left(\frac{\omega_\gamma}{M_N}\right)^L
\sqrt{\frac{L!}{(2L+1)!!}}
\left[-2(1-\lambda)\sqrt{\frac{(L+2)(L+3-2\lambda)}{2L+3}}g^E
+\sqrt{L-1+2\lambda}\,g^M\right]\,,
\end{equation}
for $J=L+1/2$, and 
\begin{equation}\label{A85}
A_\lambda(J,L)=\frac{1}{4\sqrt{\omega_\gamma}}
\left(\frac{\omega_\gamma}{M_N}\right)^{L-2}
\sqrt{\frac{L!}{(2L+1)!!}}
\left[\sqrt{\frac{(2L+1)(L-2+2\lambda)}{L-1}}g^E
-2(1-\lambda)\sqrt{L+2-2\lambda}
\left(\frac{\omega_\gamma}{M_N}\right)^2g^M\right]\,, 
\end{equation}
for $J=L-1/2$. 
The electromagnetic coupling constants calculated according to
these formulae at the resonance position
$\omega_\gamma^*=(M_{N^*}^2-M_{N}^2)/2M_{N^*}$ are given in Table~\ref{tab3}. 
For resonances with isospin $T=1/2$, they are 
split into isoscalar and isovector parts, i.e.\
\begin{equation}\label{A90}
g_{p/n}=g^{(s)}\pm g^{(v)}\,.
\end{equation}

As next we will consider the meson-baryon vertices denoting always the
meson momentum by $\vec q$. The $\pi N$ vertices can be presented in
the general form as 
\begin{equation}\label{A95}
F_{N^*(J,L)\to \pi N}=-i\frac{f_{N^*\pi N}}{m_\pi^L}
\big(\sigma_{\frac12,J}^{[L]}\cdot q^{[L]}\big)\,,
\end{equation}
and those for $\sigma N$ read 
\begin{equation}\label{A100}
F_{N^*(J,L)\to \sigma N}=\frac{f_{N^*\sigma N}}{m_\sigma^l}
\big(\sigma_{\frac12,J}^{[l]}\cdot q^{[l]}\big)\,,
\end{equation}
where $l-L\equiv 1\,\mathrm{mod}(2)$. 
For the $N^*\to \pi\Delta$ decay vertex we have 
\begin{equation}\label{A105}
F_{N^*(J,L)\to \pi\Delta}=
\sum_{{l=|J-3/2|}\atop {l\equiv L\,\mathrm{mod}(2)}}
^{J+3/2}
-i\,\frac{f_{N^*\pi\Delta}}{m_\pi^l}
\big(\sigma_{\frac32,J}^{[l]}\cdot q^{[l]}\big)\,.
\end{equation}
Because of the delta spin 3/2 the pion angular momentum $l$ in 
(\ref{A105}) 
is not fixed by the angular momentum $L$ of the resonance. 
But its possible variation is of course 
restricted by parity conservation. For the actual calculation, only in
the case of $D_{13}(1520)$ we have taken into account both, the $l=0$
and $l=2$ waves. For other resonances with $J\geq \frac32$, where the
particle data listings~\cite{PDG} do not give definite contributions
from different waves, only the lowest possible value of $l$ is taken. 

The $N^*\to\rho N$ vertex functions has a more complicated structure 
\begin{equation}\label{A110}
F_{N^*(J,L)\to\rho N}=
\sum_{j=|J-\frac12|}^{J+\frac12}\,\,\sum_{l=|j-1|}^{j+1} 
\frac{f_{N^*\rho N}}{m_\rho^l}
\big(\sigma_{\frac12,J}^{[j]}\cdot \big[q^{[l]}\otimes\epsilon_\rho^{[1]}
\big]^{[j]}\big)\,, 
\end{equation}
where the orbital momentum $l$ of the $\rho$ meson is even for $N^*$ 
with negative parity and odd for positive parity.
Again only the lowest possible values of $j$ and $l$ were used in the
calculation. 
Explicit expressions for the various vertices $F_{N^*(J,L)\to X}$ 
are listed in Table~\ref{tab2}.

Each resonance hadronic vertex contains a form factor of the form 
\begin{equation}\label{A115}
F_r(Q^2)=\frac{\Lambda_r^4}{\Lambda_r^4-(Q^2-M_r^2)^2}\,,
\end{equation}
where $Q^2$ and $M_r$ denote squared four-momentum and mass of the
resonance, respectively. For all resonances we take $\Lambda_r=1.3$ GeV. 
The same form factors were used for meson 
exchange in the $t$ channel (diagrams (2),(5)-(7), and (9)-(10) 
of Fig.~\ref{fig1}). 
For the $\pi NN$ vertex, we taken the familiar dipole form factor 
\begin{equation}\label{A120}
F_{\pi NN}(q^2)=\frac{\Lambda^2}{\Lambda^2+q^2} 
\end{equation}
with $\Lambda=0.8$ GeV. 
For the decays of $\rho$ and $\sigma$ mesons the following couplings 
were used 
\begin{equation}\label{A122}
\begin{array}{ll}
F_{\rho\to\pi\pi}^{(j=1,\mu)}=-f_{\rho\pi\pi}\epsilon_\rho^\mu\cdot
(\vec{q}_1-\vec{q}_2)\,,&
\frac{f_{\rho\pi\pi}^2}{4\pi}=\frac32\frac{m_\rho^2}{{q^*}^3}
\Gamma_{\rho\to\pi\pi}\,,\cr
F_{\sigma\to\pi\pi}^{(j=0)}=-2m_\sigma f_{\sigma\pi\pi}\,,&
\frac{f_{\sigma\pi\pi}^2}{4\pi}=\frac{1}{2q^*}\Gamma_{\sigma\to\pi\pi}\,,\cr
\end{array}
\end{equation}
with $q^*$ being the $\pi\pi$ c.m.\ momentum at $\omega_{\pi\pi}=m_\mu$, 
($\mu\in\{\rho,\sigma\}$).
Finally the resonance propagators were taken in the form 
\begin{eqnarray}\label{A124a}
G_{N^*}(W)&=&[W-M_{N^*}+\frac{i}{2}\Gamma_{N^*}(W)]^{-1}\,,\\
\label{A124b}
G_\mu(W)&=&[W^2-m_\mu^2+im_\mu\Gamma_\mu(W)]^{-1}\,, 
\quad \mu\in\{\rho,\sigma\}\,.
\end{eqnarray}

For completeness we present also the expressions for the resonance widths, in
order to fix the normalization of the hadronic coupling constants. 
For the two-body $\pi N$ decay one has
\begin{equation}\label{A125}
\Gamma_{N^*(J,L)\to\pi N}(W)=(2\pi)^4\frac{1}{2J+1}\int
\frac{d^3p}{(2\pi)^3}\frac{M_N}{E_N}\frac{d^3q}{2\omega(2\pi)^3}
\delta(W-\omega-E_N)\delta^{(3)}(\vec{q}+\vec{p}\,)
\sum\limits_{m_Jm}|\langle \frac12 m|F_{N^*\to\pi N}|Jm_J\rangle|^2\,.
\end{equation}
With the help of (\ref{A95}) one obtains 
\begin{equation}\label{A130}
\Gamma_{N^*(J,L)\to\pi N}(W)=
\frac{f_{N^*\pi N}^2}{4\pi}
\frac{2M_N}{Wm_\pi^{2L}}\frac{L!}{(2L+1)!!}q^{2L+1} 
\end{equation}
with the pion momentum in the $\pi N$ c.m.\ frame
\begin{equation}
q=\frac{\sqrt{\lambda(W,M_N,m_\pi)}}{2W}\,,
\end{equation}
where $\lambda(\alpha,\beta,\gamma)=((\alpha+\beta)^2-\gamma^2)
((\alpha-\beta)^2-\gamma^2)$.
For the decay into the $\pi\pi N$ channel we assume a sequential 
decay mechanism via an intermediate $\pi\Delta$ channel. Then one 
finds for the decay width, indicating by the superscript $(\pi\Delta)$ 
the intermediate channel in the sequential mechanism,
\begin{eqnarray}\label{A135}
\Gamma_{N^*(J,L)\to\pi\pi N}^{(\pi\Delta)}(W)&=&(2\pi)^4\frac{1}{2J+1}\int
\frac{M_N}{E_N}\frac{d^3p}{(2\pi)^3}
\frac{d^3q_1}{2\omega_1(2\pi)^3}\frac{d^3q_2}{2\omega_2(2\pi)^3}
\delta(W-\omega_1-\omega_2-E_N)\nonumber\\
&&\times\delta^{(3)}(\vec{q}_1+\vec{q}_2+\vec{p}\,)
\sum\limits_{m_Jm}|
\sum\limits_{m_\Delta}
\langle\frac12 m |F_{\Delta\to\pi N}|\frac32 m_\Delta\rangle
G_\Delta(w_\Delta)\langle\frac32 m_\Delta|F_{N^*\to\pi\Delta}|Jm_J\rangle
|^2\nonumber\\
&=&
\frac{1}{2\pi W}\frac{1}{2J+1}
\int\limits_{M_N+m_\pi}^{W-m_\pi}
dw_\Delta w_\Delta q(w_\Delta)\rho (w_\Delta)\sum
\limits_{m_Jm_\Delta}|\langle\frac32 m_\Delta|F_{N^*\to\pi\Delta}|Jm_J
\rangle|^2 \,,
\end{eqnarray}
with 
\begin{equation}\label{A140}
q(w_\Delta)=\frac{\sqrt{\lambda(W,w_\Delta,m_\pi)}}{2W}
\quad \mbox{and}\quad 
\rho(w_\Delta)=\frac{1}{2\pi}\Gamma_\Delta(w_\Delta)
|G_\Delta(w_\Delta)|^2\,,
\end{equation}
where $\Gamma_\Delta$ denotes the width of the $\Delta$ resonance. 
The final expression is formally equal to (\ref{A130})
\begin{equation}\label{A145}
\Gamma_{N^*(J,L)\to\pi\pi N}^{(\pi\Delta)}(W)=
\sum_{{l=|J-3/2|}\atop{l\equiv L\,\mathrm{mod} (2)}}^{|J+3/2|}
\frac{f^{(l)2}_{N^*\pi\Delta}}{4\pi}
\frac{2M_\Delta}{Wm_\pi^{2l}}\frac{l!}{(2l+1)!!}\overline{q}^{\,2l+1}\,,
\end{equation}
but where 
\begin{equation}\label{A150}
\overline{q}^{\,2l+1}=\int\limits_{M_N+m_\pi}^{W-m_\pi}
\frac{w_\Delta dw_\Delta}{M_\Delta}\,q(w_\Delta)^{2l+1}\rho(w_\Delta)\,.
\end{equation}
For the other type of sequential decay $N^*\to \mu N\to\pi\pi N$ 
with $\mu\in\{\rho,\sigma\}$ one has, denoting the spin of the 
meson by $j_\mu$,
\begin{eqnarray}\label{A155}
\Gamma_{N^*(J,L)\to \pi\pi N}^{(\mu N)}(W)&=&
(2\pi)^4\frac{1}{2J+1}\int
\frac{M_N}{E_N}\frac{d^3p}{(2\pi)^3}
\frac{d^3q_1}{2\omega_1(2\pi)^3}\frac{d^3q_2}{2\omega_2(2\pi)^3}
\delta(W-\omega_1-\omega_2-E_N)\nonumber\\
&&\times\delta^{(3)}(\vec{q}_1+\vec{q}_2+\vec{p}\,)
\sum\limits_{m_J,m}|
\sum\limits_{m'}
\langle Jm_J|F_{N^*\to\mu N}|\frac12 m;j_\mu m'\rangle
G_\mu(w_{\pi\pi})
F_{\mu\to\pi\pi}^{(j_\mu,m')}|^2\nonumber\\
&=&\frac{1}{2\pi}\,\frac{M_N}{W}\,\frac{1}{2J+1}
\int\limits_{2m_\pi}^{W-M_N}
dw_{\pi\pi}q(w_{\pi\pi})\rho(w_{\pi\pi})
\sum\limits_{m_J,m,m'}|\langle Jm_J|F_{N^*\to \mu N}|
\frac12m; j_\mu,m'\rangle|^2
\,.
\end{eqnarray}
with
\begin{equation}\label{A160}
q(w_{\pi\pi})=\frac{\sqrt{\lambda(W,w_{\pi\pi},M_N)}}{2W}
\quad \mbox{and}\quad 
\rho(w_{\pi\pi})=\frac{1}{2\pi}4m_\mu w_{\pi\pi}
\Gamma_\mu(w_{\pi\pi})|G_\mu(w_{\pi\pi})|^2\,.
\end{equation}
Then one has as final result 
\begin{equation}\label{A165}
\Gamma_{N^*(J,L)\to \pi\pi N}^{(\mu N)}(W)=
\sum_{{l=l_{\mathrm{min}}}\atop{l\equiv L+j_\mu+1\,\mathrm{mod}(2)}}
^{J+j_\mu+1/2}
\frac{f^{(l)2}_{N^* \mu N}}{4\pi}
\frac{2M_N}{Wm_\mu^{2l}}\frac{l!}{(2l+1)!!}\overline{q}^{2l+1}\,,
\end{equation}
where $l_{\mathrm{min}}=\mathrm{min}(|J-|j_\mu-1/2||,|J-|j_\mu+1/2||)$
and 
\begin{equation}\label{A170}
\overline{q}^{2l+1}=\int\limits_{2m_\pi}^{W-M_N}q^{2l+1}(w_{\pi\pi})
\rho(w_{\pi\pi})dw_{\pi\pi}\,.
\end{equation}

The values of the hadronic coupling constants calculated with the help of 
(\ref{A130}), (\ref{A145}), and (\ref{A165}) and partial decay widths 
from~\cite{PDG} are listed in Tables~\ref{tab3} and \ref{tab3a}.

%%%%%%%%%%%%%%%%%%%%%%%%%%%%%%%%%%%%%%%%%%%%%%%%%%%%%%%%%%%%%%%%%%%%%%%%%
\section{The amplitudes}
\label{AppC}
\renewcommand{\theequation}{C.\arabic{equation}}
\setcounter{equation}{0}

In this appendix we present detailed expressions for the amplitudes
associated with the resonance terms (diagrams (18)-(20) of Fig.~\ref{fig1})
and those background diagrams that are important in the second 
resonance region. Since the isospin
part is considered in Appendix A, we list here only the corresponding spin
structures, namely the functions 
$f^{(\pm)}(\vec{q}_1,\vec{q}_2,\vec k\,)$ appearing in
(\ref{A18}). The amplitudes $f^{(\pm)}$ 
are presented in the form of Eq.~(\ref{15}), i.e.\ by listing $K^{(\pm)}$
and $\vec L^{(\pm)}$ of the general form
\begin{equation}
f^{(\pm)}=A\left(K^{(\pm)}+i\vec{\sigma}\cdot\vec{L}^{(\pm)}\right)\,.
\end{equation}
The momenta of the participating particles are already defined in (\ref{0}). 
For convenience we introduce in addition a set of relative
momenta $\vec{p}_i$ as follows:\\
(i) The relative momentum between $\pi_1$ and the total momentum of the 
$\pi_2 N_f$ subsystem
\begin{eqnarray}\label{A180}
\vec{p}_1&=&\frac{\vec{q}_1(E_f+\omega_2)-(\vec{q}_2+\vec{p}_f)\omega_1}
{E_f+\omega_1+\omega_2}
=\vec{q}_1-\frac{\omega_1}{\omega_\gamma+E_i}(\vec{k}+\vec{p}_i)\,,
\end{eqnarray}
(ii) the relative momentum between $\pi_2$ and the final nucleon $N_f$
\begin{eqnarray}
\vec{p}_2&=&\frac{\vec{q}_2E_f-\vec{p}_f\omega_2}{E_f+\omega_2}
=\vec{q}_2-\frac{\omega_2}{E_f+\omega_2}(\vec{q}_2+\vec{p}_f)\,.
\end{eqnarray}
The argument $\omega_{\pi_i N}$ of the $\Delta$-propagator denotes the 
invariant mass of the $\pi_i N$ subsytem.
To make the formulae more compact we use the following notations for 
scalar and vector products
\begin{equation}
(ab)=(\vec{a}\cdot\vec{b})\,, \ [ab]=(\vec{a}\times\vec{b})\,.
\end{equation}
The resonance coupling constants are listed in Tables~\ref{tab3} and 
\ref{tab3a}. Other constants are given together with corresponding formulas. 
\begin{enumerate}

%---------------------------------------------------- 1
\item $\Delta$ Kroll-Ruderman (diagram (8)):
\begin{eqnarray*}
\vec{L}^{(\pm)}&=&[p_2\epsilon]G_\Delta(w_{\pi_2N})
\pm(1\leftrightarrow 2)\,,\\ 
K^{(\pm)}&=&-(p_2\epsilon)G_\Delta(w_{\pi_2N})\pm(1\leftrightarrow 2)\,,\\ 
A&=&\frac{e}{3}
\left(\frac{f_{\Delta\pi N}}{m_\pi}\right)^2\,. 
\end{eqnarray*}

%---------------------------------------------------- 2
\item $N\Delta$ u-channel term (the second diagram from the group (12),
M1 $N\to\Delta$ transition):
\begin{eqnarray*}
\vec{L}^{(\pm)}&=&\Big[
2(q_1p_2)[k\epsilon]-(p_2\epsilon)[q_1k]+(p_2k)[q_1\epsilon]
+2(q_1\epsilon)[p_2k]\\
&&-2(q_1k)[p_2\epsilon]
\Big]G_\Delta(w_{\pi_2N})
\Big(E_i-\omega_1-\sqrt{(\vec{p}_i-\vec{q}_1)^2+M_N^2}\Big)^{-1}
\pm(1\leftrightarrow 2)
\,,\\ 
K^{(\pm)}&=&\Big[(q_1k)(p_2\epsilon)-(p_2k)(q_1\epsilon)\Big]
G_\Delta(w_{\pi_2N})
\Big(E_i-\omega_1-\sqrt{(\vec{p}_i-\vec{q}_1)^2+M_N^2}\Big)^{-1}
\pm(1\leftrightarrow 2)
\,,\\ 
A&=&-\frac{g^M\,f_{\Delta\pi N}f_{\pi NN}}{12\sqrt{2}M_N m_\pi^2}\,, \\ 
g^M&=&-1.845,\ f_{\pi NN}=1.0
\end{eqnarray*}

%---------------------------------------------------- 3
\item $\Delta\Delta$ u-channel term (the second diagram from the group (14),  
M1 $\Delta\to\Delta$ transition):
\begin{eqnarray*}
\vec{L}^{(\pm)}&=&\Big[
2(q_1p_2)[k\epsilon]-(p_2\epsilon)[q_1k]+(p_2k)[q_1\epsilon]
-(q_1\epsilon)[p_2k]\\
&&+(q_1k)[p_2\epsilon]\Big]G_\Delta(w_{\pi_2N})
\Big(E_i-\omega_1-\sqrt{(\vec{p}_i-\vec{q}_1)^2+M_\Delta^2}\Big)^{-1}
\pm(1\leftrightarrow 2)\,,\\ 
K^{(\pm)}&=&5\Big[
-(q_1k)(p_2\epsilon)+(p_2k)(q_1\epsilon)\Big]G_\Delta(w_{\pi_2N})
\Big(E_i-\omega_1-\sqrt{(\vec{p}_i-\vec{q}_1)^2+M_\Delta^2}\Big)^{-1}
\pm(1\leftrightarrow 2)
\,,\\ 
A&=&\frac{e\mu_p}{36M_N}\left(\frac{f_{\Delta\pi N}}{m_\pi}\right)^2
\,,\\ \mu_p&=&2.79 
\end{eqnarray*}

%---------------------------------------------------- 4
\item $\rho$ Kroll-Ruderman term (diagram(4)):
\begin{eqnarray*}
\vec{L}^{(-)}&=&[q_2\epsilon]-[q_1\epsilon]\,,\quad \vec{L}^{(+)}=0
\,,\\
K^{(\pm)}&=&0\,,\\
A&=&e(1+\kappa_\rho)\frac{f_{\rho NN}}{2M_N}f_{\rho\pi\pi}
G_\rho(w_{\pi\pi})\,, \\ f_{\rho NN}&=&2.24\,,\quad
\kappa_\rho=6\,,\quad f_{\rho \pi\pi}=6.02\,.
\end{eqnarray*}

%---------------------------------------------------- 5
\item $\rho$ photoproduction via $\sigma$ exchange (diagram(6)):
\begin{eqnarray*}
\vec{L}^{(\pm)}&=&0\,,\\
K^{(-)}&=&(q_1\epsilon)\omega_\gamma(\omega_1+\omega_2)
+2(q_2\epsilon)(q_1k)-(1\leftrightarrow 2)\,,\\
K^{(+)}&=&0\,,\\
A&=&-e\frac{f_{\gamma\rho\sigma}}{m_\rho}\frac{f_{\sigma NN}f_{\rho\pi\pi}}
{t-m_\sigma^2}G_\rho(w_{\pi\pi})\,, \\ 
f_{\gamma\rho\sigma}&=&2.2\,,\quad f_{\sigma NN}=10.02\,.
\end{eqnarray*}

%---------------------------------------------------- 6
\item $\gamma(E1)N\to N^*\left(\frac12,0\right)\to\pi\Delta$:
\begin{eqnarray*}
\vec{L}^{(\pm)}
&=&\Big[(p_1p_2)[p_1\epsilon]
-\frac13 p_1^2[p_2\epsilon]\Big]G_\Delta(w_{\pi_2N})
\pm(1\leftrightarrow 2)\,,\\
K^{(\pm)}&=&\Big[(p_1p_2)(p_1\epsilon)
-\frac13 p_1^2(p_2\epsilon)\Big]G_\Delta(w_{\pi_2N})
\pm(1\leftrightarrow 2)
\,,\\ 
A&=&-\frac{f_{N^*\pi\Delta}f_{\Delta\pi N}}{2\sqrt{30}m_\pi^3}
G_{N^*}(W)\,. 
\end{eqnarray*}

%---------------------------------------------------- 7
\item $\gamma(E1)N\to N^*\left(\frac12,0\right)\to\rho N$:
\begin{eqnarray*}
\vec{L}^{(-)}&=&[p_1\epsilon]-[p_2\epsilon]\,,\quad \vec{L}^{(+)}=0\,,\\
K^{(-)}&=&(p_1\epsilon)-(p_2\epsilon)\,,\quad K^{(+)}=0\,,\\ 
A&=&-\frac{1}{6}
f_{N^*\rho N}f_{\rho\pi\pi}G_{N^*}(W)G_\rho(w_{\pi\pi})\,.
\end{eqnarray*}

%---------------------------------------------------- 8
\item $\gamma(E1)N\to N^*\left(\frac12,0\right)\to\sigma N$:
\begin{eqnarray*}
\vec{L}^{(+)}&=&[p_1\epsilon]+[p_2\epsilon]\,,\quad \vec{L}^{(-)}=0\,,\\
K^{(+)}&=&(p_1\epsilon)+(p_2\epsilon)\,,\quad K^{(-)}=0\,,\\
A&=&-\frac{1}{3}
f_{N^*\sigma N}f_{\sigma\pi\pi}G_{N^*}(W)G_\sigma(w_{\pi\pi})\,. 
\end{eqnarray*}

%---------------------------------------------------- 9
\item $\gamma(M1)N\to N^*\left(\frac12,1\right)\to \pi\Delta$:
\begin{eqnarray*}
\vec{L}^{(-)}&=&\Big[
(p_2k)[p_1\epsilon]+(p_1\epsilon)[p_2k]\Big]G_\Delta(w_{\pi_2N})
-(1\leftrightarrow 2)\,, \\
\vec{L}^{(+)}&=&(p_1p_2)[k\epsilon]G_\Delta(w_{\pi_2N})
+(1\leftrightarrow 2)\,,\\
K^{(-)}&=&(p_2k)(p_1\epsilon)G_\Delta(w_{\pi_2N})
-(1\leftrightarrow 2)\,,\quad K^{(+)}=0\,,\\
A&=&-\frac{f_{N^*\pi\Delta}f_{\Delta\pi N}}{6\sqrt{6}M_Nm_\pi^2}
G_{N^*}(W)\,. 
\end{eqnarray*}

%---------------------------------------------------- 10
\item $\gamma(M1)N\to N^*\left(\frac12,1\right)\to\sigma N$:
\begin{eqnarray*}
\vec{L}^{(+)}&=&[k\epsilon]\,,\quad \vec{L}^{(-)}=0\,,\\
K^{(\pm)}&=&0\,,\\
A&=&-\frac{m_\sigma f_{N^*\sigma N}f_{\sigma\pi\pi}}{\sqrt{6}M_N}
G_{N^*}(W)G_\sigma(\omega_{\pi\pi})\,. 
\end{eqnarray*}

%---------------------------------------------------- 11
\item $\gamma(E1) N\to N^*\left(\frac32,2\right)\to \pi\Delta$:
\begin{eqnarray*}
\vec{L}^{(\pm)}&=&\Big[
-\frac{f_{N^*\pi\Delta}^{(s)}}{2}[p_2\epsilon]
-\sqrt{\frac{6}{5}}
\frac{f_{N^*\pi\Delta}^{(d)}}{2m_\pi^2}\big(
[p_1p_2](p_1\epsilon)-(p_1p_2)[p_1\epsilon]
+\frac{2}{3}p_1^2[p_2\epsilon]\big)\Big]G_\Delta(w_{\pi_2N})
\pm(1\leftrightarrow 2)
\,,\\
K^{(\pm)}&=&\Big[
f_{N^*\pi\Delta}^{(s)}(p_2\epsilon)
-\sqrt{\frac{6}{5}}\frac{f_{N^*\pi\Delta}^{(d)}}{2m_\pi^2}\big(
(p_1p_2)(p_1\epsilon)-\frac{1}{3}p_1^2(p_2\epsilon)\big)\Big]
G_\Delta(w_{\pi_2N})\pm(1\leftrightarrow 2)\,,\\
A&=&-\frac{f_{\Delta\pi N}}{6m_\pi}G_{N^*}(W)\,. 
\end{eqnarray*}

%---------------------------------------------------- 12
\item $\gamma(M2) N\to N^*\left(\frac32,2\right)\to \pi\Delta$:
\begin{eqnarray*}
\vec{L}^{(\pm)}&=&\Big[
\frac{f_{N^*\pi\Delta}^{(s)}}{2}
\big(2(p_2k)[k\epsilon]-k^2[p_2\epsilon]\big)\\
&&-\sqrt{\frac{6}{5}}\frac{f_{N^*\pi\Delta}^{(d)}}{6m_\pi^2}\big( 
[k\epsilon]\{3(p_1k)(p_1p_2)-p_1^2(p_2k)\}
+[p_2k](p_1k)(p_1\epsilon)\\
&&+[p_1\epsilon]\{(p_1k)(p_2k)-2(p_1p_2)k^2\}
+[p_2\epsilon]\{p_1^2k^2-(p_1k)^2\}\\
&&-[p_1k]\{2(p_1k)(p_2\epsilon)-(p_1\epsilon)(p_2k)\}\big)\Big]
G_\Delta(w_{\pi_2N}) 
\pm (1\leftrightarrow 2)
\,,\\
K^{(\pm)}&=&\sqrt{\frac{6}{5}}\frac{f_{N^*\pi\Delta}^{(d)}}{6m_\pi^2}\Big[
2(p_1k)^2(p_2\epsilon)-2(p_1k)(p_1\epsilon)(p_2k)
-k^2p_1^2(p_2\epsilon)+k^2(p_1\epsilon)(p_1p_2)\Big]G_\Delta(w_{\pi_2N}) 
\pm (1\leftrightarrow 2)
\,,\\
A&=&-\frac{f_{\Delta\pi N}}{2\sqrt{15}M_N^2 m_\pi}G_{N^*}(W)\,. 
\end{eqnarray*}

%---------------------------------------------------- 13
\item $\gamma(E1)N\to N^*\left(\frac32,2\right)\to \rho N$:
\begin{eqnarray*}
\vec{L}^{(-)}&=&[p_2\epsilon]-[p_1\epsilon]\,,\quad \vec{L}^{(+)}=0\,,\\
K^{(-)}&=&2(p_1\epsilon)-2(p_2\epsilon)\,,\quad K^{(+)}=0\,,\\ 
A&=&-\frac{1}{6}f_{N^*\rho N}f_{\rho \pi\pi}G_{N^*}(W)G_\rho(w_{\pi\pi})\,. 
\end{eqnarray*}

%---------------------------------------------------- 14
\item $\gamma(M2)N\to N\left(\frac32,2\right)\to \rho N$:
\begin{eqnarray*}
\vec{L}^{(-)}&=&2[k\epsilon](p_1k)
-k^2\big[p_1\epsilon]-(1\leftrightarrow 2)\,,\quad \vec{L}^{(+)}=0\,,\\
K^{(\pm)}&=&0\,,\\ 
A&=&-\frac{f_{N^*\rho N}f_{\rho \pi\pi}}{2\sqrt{15}M_N^2}
G_{N^*}(W)G_\rho(w_{\pi\pi})\,.
\end{eqnarray*}

%---------------------------------------------------- 15
\item $\gamma(M1)N\to N\left(\frac32,1\right)\to\pi\Delta$:
\begin{eqnarray*}
\vec{L}^{(-)}&=&\frac52\Big[
(p_2k)[p_1\epsilon]-(p_2\epsilon)[p_1k]\Big]G_\Delta(w_{\pi_2N})
-(1\leftrightarrow 2)\,,\\
\vec{L}^{(+)}&=&\Big[
(p_1p_2)[k\epsilon]+
\frac32\big((p_2\epsilon)[p_1k]-(p_2k)[p_1\epsilon]\big)\Big]
G_\Delta(w_{\pi_2N})+(1\leftrightarrow 2)\,,\\
K^{(-)}&=&5(p_1k)(p_2\epsilon)G_\Delta(w_{\pi_2N})
-(1\leftrightarrow 2)\,,\quad
K^{(+)}=0\,,\\
A&=&-\frac{f_{N^*\pi\Delta}f_{\Delta\pi N}}
{6\sqrt{30}M_Nm_\pi^2}G_{N^*}(W)\,. 
\end{eqnarray*}

%---------------------------------------------------- 16
\item $\gamma(E2)N\to N^*\left(\frac32,1\right)\to\rho N$:
\begin{eqnarray*}
\vec{L}^{(-)}&=&(p_1\epsilon)[p_2k]+(p_1k)[p_2\epsilon]
-(1\leftrightarrow 2)\,,\quad \vec{L}^{(+)}=0
\,,\\
K^{(\pm)}&=&0\,,\\
A&=&\frac{f_{N^*\rho N}f_{\rho\pi\pi}}{2\sqrt{5}m_\rho M_N}
G_{N^*}(W)G_\rho(w_{\pi\pi})\,. 
\end{eqnarray*}

%---------------------------------------------------- 17
\item $\gamma(M1)N\to N^*\left(\frac32,2\right)\to\rho N$:
\begin{eqnarray*}
\vec{L}^{(-)}&=&(p_2k)[p_1\epsilon]+(p_1\epsilon)[p_2k]
-(1\leftrightarrow 2)\,,\quad \vec{L}^{(+)}=0\,,\\
K^{(-)}&=&2(p_1k)(p_2\epsilon)-(1\leftrightarrow 2)
\,,\quad K^{(+)}=0\,,\\
A&=&-\frac{f_{N^*\rho N}f_{\rho\pi\pi}}{6m_\rho M_N}
G_{N^*}(W)G_\rho(w_{\pi\pi})\,.
\end{eqnarray*}

%---------------------------------------------------- 18
\item $\gamma(E3)N\to N^*\left(\frac52,2\right)\to\pi\Delta$:
\begin{eqnarray*}
\vec{L}^{(\pm)}&=&\Big[
[p_1\epsilon]\big(-10k^2(p_1p_2)+22(p_1k)(p_2k)\big)
+[p_1k]\big(8(p_1\epsilon)(p_2k)+15(p_2\epsilon)(p_1k)\big)\\
&&+[p_2\epsilon]\big(9p_1^2k^2-17(p_1k)^2\big)-13[p_2k](p_1\epsilon)(p_1k)
+7[k\epsilon]\big((p_1p_2)(p_1k)\\
&&-p_1^2(p_2k)\big)\Big]G_\Delta(w_{\pi_2N})
\pm (1\leftrightarrow 2)
\,,\\
K^{(\pm)}&=&\Big[
2(p_1\epsilon)\big(5(p_1k)(p_2k)-(p_1p_2)k^2\big)
+(p_2\epsilon)\big(5(p_1k)^2-p_1^2k^2\big)\Big]G_\Delta(w_{\pi_2N})
\pm (1\leftrightarrow 2)
\,,\\ 
A&=&-\frac{\sqrt{2}}{105}
\frac{f_{N^*\pi\Delta}f_{\Delta\pi N}}{4M_N^2m_\pi^3}
G_{N^*}(W)\,.
\end{eqnarray*}

%---------------------------------------------------- 19
\item $\gamma(M2)N\to N^*\left(\frac52,2\right)\to\pi\Delta$:
\begin{eqnarray*}
\vec{L}^{(\pm)}&=&\frac{1}{7}\Big[
-[p_1\epsilon]\big(k^2(p_1p_2)+2(p_1k)(p_2k)\big)
+2[p_1k]\big(4(p_1\epsilon)(p_2k)-3(p_2\epsilon)(p_1k)\big)\\
&&+[p_2\epsilon]\big(3p_1^2k^2-8(p_1k)^2\big)
+8[p_2k](p_1\epsilon)(p_1k)
+2[k\epsilon]\big(2(p_1p_2)(p_1k)+p_1^2(p_2k)\big)\Big]G_\Delta(w_{\pi_2N})
\pm (1\leftrightarrow 2)
\,,\\
K^{(\pm)}&=&\Big[
(p_1\epsilon)\big(k^2(p_1p_2)-2(p_1k)(p_2k)\big)
+(p_2\epsilon)\big(2(p_1k)^2-p_1^2k^2\big)\Big]G_\Delta(w_{\pi_2N})
\pm (1\leftrightarrow 2)
\,,\\ 
A&=&-\frac{\sqrt{7}}{15}
\frac{f_{N^*\pi\Delta}f_{\Delta\pi N}}{4M_N^2m_\pi^3}
G_{N^*}(W)\,.
\end{eqnarray*}

%---------------------------------------------------- 20
\item $\gamma(E2)N\to N^*\left(\frac52,3\right)\to\pi\Delta$:
\begin{eqnarray*}
\vec{L}^{(+)}&=&\Big[
-(p_2k)[p_1\epsilon]-(p_2\epsilon)[p_1k]\Big]G_\Delta(w_{\pi_2N})
+(1\leftrightarrow 2)\,,\quad \vec{L}^{(-)}=0\,,\\
K^{(+)}&=&3(p_2k)(p_1\epsilon)G_\Delta(w_{\pi_2N})+(1\leftrightarrow 2)\,,
\quad K^{(-)}=0\,,\\ 
A&=&-\frac{1}{10}
\frac{f_{N^*\pi\Delta}f_{\Delta\pi N}}{2M_Nm_\pi^2}
G_{N^*}(W)\,.
\end{eqnarray*}

%---------------------------------------------------- 21
\item $\gamma(M3)N\to N\left(\frac52,3\right)\to\pi\Delta$:
\begin{eqnarray*}
\vec{L}^{(+)}&=&\Big[
k^2\big(4(p_2k)[p_1\epsilon] 
+(p_2\epsilon)[p_1k]+\frac32[k\epsilon](p_1p_2)\big)
-\frac{15}{2}[k\epsilon](p_1k)(p_2k)\Big]G_\Delta(w_{\pi_2N})
+(1\leftrightarrow 2)\,,\\
\vec{L}^{(-)}&=&0\,,\\
K^{(\pm)}&=&0\,,\\
A&=&\frac{1}{10\sqrt{7}}
\frac{f_{N^*\pi\Delta}f_{\Delta\pi N}}{2M_N^3m_\pi^2}
G_{N^*}(W)\,.
\end{eqnarray*}

%---------------------------------------------------- 22
\item $\gamma(E2)N\to N\left(\frac52,3\right)\to\rho N$:
\begin{eqnarray*}
\vec{L}^{(-)}&=&(p_1k)[p_1\epsilon]+(p_1\epsilon)[p_1k] 
-(1\leftrightarrow 2)\,,\quad \vec{L}^{(+)}=0\,,\\
K^{(-)}&=&3(p_2\epsilon)(p_2k)-(1\leftrightarrow 2)
\,,\quad K^{(+)}=0\,,\\
A&=&\frac{1}{5}\frac{f_{N^*\rho N}f_{\rho\pi\pi}}{2M_Nm_\rho}
G_{N^*}(W)G_\rho(w_{\pi\pi})\,.
\end{eqnarray*}

%---------------------------------------------------- 23
\item $\gamma(M3)N\to N\left(\frac52,3\right)\to\rho N$:
\begin{eqnarray*}
\vec{L}^{(-)}&=&k^2\big(4(p_1k)[p_1\epsilon]
+(p_1\epsilon)[p_1k]+\frac{3}{4}(p_1^2-p_2^2)
[k\epsilon]\big)-\frac{15}{4}[k\epsilon]\big((p_1k)^2-(p_2k)^2\big)
-(1\leftrightarrow 2)\,,\\
\vec{L}^{(+)}&=&0\,,\\
K^{(\pm)}&=&0\,,\\
A&=&\frac{1}{5\sqrt{7}}\frac{f_{N^*\rho N}f_{\rho\pi\pi}}
{2M_N^3m_\rho}G_{N^*}(W)G_\rho(w_{\pi\pi})\,.\\ 
\end{eqnarray*}

%---------------------------------------------------- 24
\item $\gamma(E2)N\to N\left(\frac52,3\right)\to\sigma N$:
\begin{eqnarray*}
\vec{L}^{(+)}&=&-\big((p_1k)+(p_2k)\big)\big([p_1\epsilon]+[p_2\epsilon]
\big)-\big((p_1\epsilon)+(p_2\epsilon)\big)\big([p_1k]+[p_2k]\big)\,,
\quad \vec{L}^{(-)}=0\,,\\
K^{(+)}&=&3\big[(p_1\epsilon)+(p_2\epsilon)\big]\big[(p_1k)+(p_2k)\big]\,,
\quad K^{(-)}=0\,,\\
A&=&-\frac{1}{5}\frac{f_{N^*\sigma N}f_{\sigma\pi\pi}}
{M_Nm_\sigma}G_{N^*}(W)G_\sigma(w_{\pi\pi})\,. 
\end{eqnarray*}

%---------------------------------------------------- 25
\item $\gamma(M3)N\to N\left(\frac52,3\right)\to\sigma N$:
\begin{eqnarray*}
\vec{L}^{(+)}&=&k^2\Big[4((p_1k)+(p_2k))([p_1\epsilon]+[p_2\epsilon])
+([p_1k]+[p_2k])((p_1\epsilon)+(p_2\epsilon))\\
&&+\frac{3}{2}(\vec{p}_1+\vec{p}_2)^2[k\epsilon]\Big]
-\frac{15}{2}[k\epsilon]\Big[(p_1k)+(p_2k)\Big]^2\,,\quad \vec{L}^{(-)}=0\,,\\
K^{(\pm)}&=&0\,,\\
A&=&\frac{1}{5\sqrt{7}}\frac{f_{N^*\sigma N}f_{\sigma\pi\pi}}
{M_N^3m_\sigma}G_{N^*}(W)G_\sigma(w_{\pi\pi})\,.
\end{eqnarray*}

\end{enumerate}

%% file: pap3_tab.tex
%%%%%%%%%%%%%%%%%%%%%%%%%%%%%%%%%%%%%%%%%%%%%%%%%%%%%%%%%%%%%%%%%%%%%%%%%
\begin{table}
\renewcommand{\arraystretch}{2.0}
\caption{\label{tab1}
Listing of resonances included in the model. The partial decay widths are
given in percent. The two values of $\Gamma_{\pi\Delta}/\Gamma$ 
for $D_{13}\to\pi\Delta$ and 
$D_{33}\to\pi\Delta$ correspond to $s$- and $d$-wave $\pi\Delta$ states. 
For $S_{11}(1535)$ the partial $\eta N$ decay width is equal to the 
$\pi N$ width.}  
\begin{ruledtabular}
\begin{tabular}{c|ccccccccc}
$L_{2T2J}(M^*)\ \ $ & M [MeV] & $\Gamma$ [MeV] & $\Gamma_{\pi N}/\Gamma$ 
& $\Gamma_{\pi\Delta}/\Gamma$ & $\Gamma_{\rho N}/\Gamma$  
& $\Gamma_{\sigma N}/\Gamma$ & Multipoles \\
\colrule  
$P_{33}(1232)$ & 1232 & 120  & 100 &  &  &  & E2, M1 \\
$P_{11}(1440)$ & 1440 & 350 
 & 67 & 25 &  & 8 & M1 \\
$D_{13}(1520)$ & 1520 & 120 
 & 59 & 9, 12 & 20 &  & E1, M2 \\
$S_{11}(1535)$ & 1535 & 150 
 & 45 &    & 5 & 5 & E1 \\
$S_{31}(1620)$ & 1620 & 150
 & 25 & 55 & 20 & & E1 \\
$D_{15}(1675)$ & 1675 & 150
 & 45 & 55 &  &  & E3, M2 \\
$F_{15}(1680)$ & 1680 & 130
 & 69 & 10 & 9 & 12 & E2, M3 \\
$D_{33}(1700)$ & 1700 & 300
 & 15 & 41, 4 & 40 &  & E1, M2  \\
$P_{13}(1720)$ & 1720 & 150 
 & 15 &  & 85 &  & E2, M1  \\
\end{tabular}
\end{ruledtabular}\end{table}
%%%%%%%%%%%%%%%%%%%%%%%%%%%%%%%%%%%%%%%%%%%%%%%%%%%%%%%%%%%%%%%%%%%%%%%%%
\begin{table}[ht]
\renewcommand{\arraystretch}{1.5}
\caption{\label{tab3}Listing of coupling constants for \protect$T=1/2$ 
resonances. The signs of the hadronic constants are chosen according to 
$\pi$ and $\pi\pi$ production analyses as explained in Sect.~\ref{gammapipi} 
in the paragraph following Eq.~(\ref{25}). The two values for 
$f_{N^*\pi\Delta}$ for $D_{13}(1520)$ refer to $s$- and $d$-wave $\pi\Delta$ 
states.
} 
\begin{ruledtabular} 
\begin{tabular}{c|dddddddd}
$L_{2T2J}(M^*)$ & \multicolumn{1}{c}{$g^{E(s)}$} & 
\multicolumn{1}{c}{$g^{E(v)}$} & \multicolumn{1}{c}{$g^{M(s)}$} & 
\multicolumn{1}{c}{$g^{M(v)}$}& \multicolumn{1}{c}{$f_{N^*\pi N}$} & 
\multicolumn{1}{c}{$f_{N^*\pi\Delta}$} & \multicolumn{1}{c}{$f_{N^*\rho N}$} 
& \multicolumn{1}{c}{$f_{N^*\sigma N}$} \\
\colrule 
$P_{11}(1440)$ &  &  & 0.089 & 0.375 & -1.454 & -4.219 & & -3.187 \\
$D_{13}(1520)$ & -0.015 & 0.236 & 0.640 & 0.915 & -0.323 & 
\multicolumn{1}{c}{0.791, 0.846}& 7.651 &  \\
$S_{11}(1535)$ & -0.053 & -0.163 & & & -1.219 &  & 3.842 & -3.851 \\
$D_{15}(1670)$ & -0.032 & -0.052 & -0.227 & 0.443 & -0.196 
& -0.706 & & \\
$F_{15}(1680)$ & 0.202 & 0.248 & 0.510 & 1.452 & -0.082 & 0.458 & 7.888 
& -6.572 \\
$P_{13}(1720)$ & -0.109 & -0.026 & -0.067 & 0.036 & 0.269 & & 18.840 & \\
\end{tabular}
\end{ruledtabular} 
\end{table}
%%%%%%%%%%%%%%%%%%%%%%%%%%%%%%%%%%%%%%%%%%%%%%%%%%%%%%%%%%%%%%%%%%%%%%%%%
\begin{table}[ht]
\renewcommand{\arraystretch}{2.0}
\caption{\label{tab3a}Same as in Table~\protect\ref{tab3} for \protect$T=3/2$ 
resonances.} 
\begin{ruledtabular}
\begin{tabular}{c|ddddd}
$L_{2T2J}(M^*)\ \ $ & \multicolumn{1}{c}{$g^E$} & \multicolumn{1}{c}{$g^M$}
 & \multicolumn{1}{c}{$f_{N^*\pi N}$} & \multicolumn{1}{c}{$f_{N^*\pi\Delta}$}
 & \multicolumn{1}{c}{$f_{N^*\rho N}$} \\
\colrule   
$P_{33}(1232)$ & -0.087 & -1.845 & 2.230 & 2.982 & \\
$S_{31}(1620)$ & -0.069 & & 0.879 & 1.068 & -4.067 \\
$D_{33}(1700)$ &  0.236 & -0.506 & 0.149 & \multicolumn{1}{c}{2.008, 0.237} 
& 4.417\\
\end{tabular}
\end{ruledtabular}
\end{table}
%%%%%%%%%%%%%%%%%%%%%%%%%%%%%%%%%%%%%%%%%%%%%%%%%%%%%%%%%%%%%%%%%%%%%%%%%
\begin{table}[ht]
\renewcommand{\arraystretch}{2.5}
\caption{\label{tab2}Listing of hadronic vertices $F_{N^*(J,L)\to X}$. The 
empty spaces in the last two columns indicate that the corresponding couplings
are insignificant or not provided by the PDG compilation~\protect\cite{PDG}
for the resonances of the model.} 
\begin{ruledtabular}
\begin{tabular}{c|cccc}
$J,L\ \ $ & $X=\pi N$ & $\pi\Delta$ & $\rho N$ & $\sigma N$ \\
\colrule          %\hline\hline
$\frac12,\,0$ & $-i\,f_{N^*\pi N}$ & 
$-i\,\frac{f_{N^*\pi\Delta}}{m_\pi^2}\big(\sigma_{\frac32,\frac12}^{[2]}\cdot
q^{[2]}\big)$ & $f_{N^*\rho N}
\big(\sigma_{\frac12,\frac12}^{[1]}\cdot\epsilon_\rho^{[1]}\big)$ &
$\frac{f_{N^*\sigma N}}{m_\sigma}
\big(\sigma_{\frac12,\frac12}^{[1]}\cdot q^{[1]}\big)$ \\
$\frac12,\,1$ & $-i\,\frac{f_{N^*\pi N}}{m_\pi}
\big(\sigma_{\frac12,\frac12}^{[1]}\cdot q^{[1]}\big)$ & 
$-i\,\frac{f_{N^*\pi\Delta}}{m_\pi}
\big(\sigma_{\frac32,\frac12}^{[1]}\cdot q^{[1]}\big)$ & & 
$f_{N^*\sigma N}$ \\
$\frac32,\,1$ & $-i\,\frac{f_{N^*\pi N}}{m_\pi}
\big(\sigma_{\frac12,\frac32}^{[1]}\cdot q^{[1]}\big)$ & 
$-i\,\frac{f_{N^*\pi\Delta}}{m_\pi}
\big(\sigma_{\frac32,\frac32}^{[1]}\cdot q^{[1]}\big)$ & 
$\frac{f_{N^*\rho N}}{m_\rho}
\big(\sigma_{\frac12,\frac32}^{[1]}\cdot[q^{[1]}\otimes
\epsilon_\rho^{[1]}]^{[1]}
\big)$ & \\
$\frac32,\,2$ & $-i\,\frac{f_{N^*\pi N}}{m_\pi^2}
\big(\sigma_{\frac12,\frac32}^{[2]}\cdot q^{[2]}\big)$
& $ -i\,f_{N^*\pi\Delta}^{(s)}
-i\,\frac{f_{N^*\pi\Delta}^{(d)}}{m_\pi^2}
\big(\sigma_{\frac32,\frac32}^{[2]}\cdot q^{[2]}\big)$ & 
$f_{N^*\rho N}\big(\sigma_{\frac12,\frac32}^{[1]}
\cdot\epsilon_\rho^{[1]}\big)$& \\
$\frac52,\,2$ & $-i\,\frac{f_{N^*\pi N}}{m_\pi^2}
\big(\sigma_{\frac12,\frac52}^{[2]}\cdot q^{[2]}\big)$ & 
$-i\,\frac{f_{N^*\pi\Delta}}{m_\pi^2}
\big(\sigma_{\frac32,\frac52}^{[2]}\cdot q^{[2]}\big)$ & & \\
$\frac52,\,3$ & $-i\,\frac{f_{N^*\pi N}}{m_\pi^3}
\big(\sigma_{\frac12,\frac52}^{[3]}\cdot q^{[3]}\big)$ & 
$-i\,\frac{f_{N^*\pi\Delta}}{m_\pi}
\big(\sigma_{\frac32,\frac52}^{[1]}\cdot q^{[1]}\big)$ & 
$\frac{f_{N^*\rho N}}{m_\rho}
\big(\sigma_{\frac12,\frac52}^{[2]}\cdot
[q^{[1]}\otimes\epsilon_\rho^{[1]}]^{[2]}\big)$ & 
$\frac{f_{N^*\sigma N}}{m_\sigma^2}
\big(\sigma_{\frac12,\frac52}^{[2]}\cdot q^{[2]}\big)$ \\
%& $\frac{2}{35}(\frac{f_{N^*\pi N}}{m_\pi^3})^2q^6 $ &
%$\frac{1}{3}(\frac{f_{N^*\pi\Delta}}{m_\pi})^2q^2 $ &
%$\frac{8}{15}(\frac{f_{N^*\rho N}}{m_\rho})^2q^2 $ & 
%$\frac{1}{5}(\frac{f_{N^*\sigma N}}{m_\sigma^2})^2q^4 $ 
\end{tabular}
\end{ruledtabular}
\end{table}
%%%%%%%%%%%%%%%%%%%%%%%%%%%%%%%%%%%%%%%%%%%%%%%%%%%%%%%%%%%%%%%%%%%%%%%%%
\begin{table}[ht]
\renewcommand{\arraystretch}{2.0}
\caption{\label{tab4}Isospin coefficients $C_l^{(\pm)}$ (\ref{A18}) 
for the resonance amplitudes. The isospin of a resonance is indicated 
in parenthesis. The coefficients contain electromagnetic coupling constants 
$g$ introduced in~(\protect\ref{A65a}). For the resonances with isospin 1/2 
they are split into isoscalar $g^{(s)}$ and isovector $g^{(v)}$ parts.
In the lower part we give the corresponding coefficients for several 
important Born amplitudes where the associated diagram is indicated
in parenthesis. 
%$\Delta$ Kroll-Ruderman term (diagram (8)), $N\Delta$ $u$-channel transition 
%(second diagram from the group (12)), $\Delta\Delta$ $u$-channel transition 
%(second diagram from (14)), $\rho$ Kroll-Ruderman term (diagram (4)), and  
%$\rho$-photoproduction via $\sigma$-exchange (diagram (6)).
} 
\begin{ruledtabular}
\begin{tabular}{c|cccccc}
 &$C^{(+)}_1$&$C^{(+)}_2$&$C^{(+)}_3$&$C^{(-)}_1$&$C^{(-)}_2$&$C^{(-)}_3$ \\
\colrule          %\hline\hline
$\gamma N\to N^*(1/2)\to\pi\Delta\ \ $ & $-\frac{2\sqrt{2}}{3}g^{(s)}$ 
& $-\frac{2\sqrt{2}}{3}g^{(v)}$ & & $\frac{\sqrt{2}}{3}g^{(s)}$ 
& $-\frac{\sqrt{2}}{3}g^{(v)}$ & $-\frac{2\sqrt{2}}{3}g^{(v)}$ \\
$\gamma N\to N^*(1/2)\to\rho N$ & & & & $\frac{1}{\sqrt{3}}g^{(s)}$
& $-\frac{1}{\sqrt{3}}g^{(v)}$ & $-\frac{2}{\sqrt{3}}g^{(v)}$ \\
$\gamma N\to N^*(1/2)\to\sigma N$  & $-\sqrt{\frac23}g^{(s)}$ &
$-\sqrt{\frac23}g^{(v)}$ & & & & \\
$\gamma N\to N^*(3/2)\to\pi\Delta$ & & $-\frac{2}{3\sqrt{10}}g$ 
& $\frac{2}{\sqrt{10}}g$
& & $-\frac{\sqrt{10}}{3}g$ & $\frac{\sqrt{10}}{3}g$ \\
$\gamma N\to N^*(3/2)\to\rho N$ & & & & & $-\sqrt{\frac23}g$ 
& $\sqrt{\frac23}g$ \\
\colrule          %\hline\hline
$\Delta-$KR (8)& & $-\frac23$ & $\frac23$ & & $-\frac43$ & $\frac23$ \\
$N\Delta(u\mbox{-channel})$ (2nd of 12) 
& & $\sqrt{\frac23}$ & $\sqrt{\frac23}$ & & $-\sqrt{\frac23}$ 
& $\sqrt{6}$ \\
$\Delta\Delta(u\mbox{-channel})$ (2nd of 14)
& $\frac23$ & $\frac43$ & $-\frac23$ & $-\frac13$ & $\frac53$ & \\
$\rho-$KR (4)& & & & & & $-2$\\
$\gamma\to\rho$ ($\sigma$-exchange) (6)& & & & & $-1$& \\
\end{tabular}
\end{ruledtabular}
\end{table}
%%%%%%%%%%%%%%%%%%%%%%%%%%%%%%%%%%%%%%%%%%%%%%%%%%%%%%%%%%%%%%%%%%%%%%%%%
\begin{table}[ht]
\renewcommand{\arraystretch}{1.0}
\caption{\label{tabGDH} Contribution of individual charge channels of 
double pion photoproduction on nucleon and deuteron to the finite GDH 
integral (in $\mu$b), evaluated up to 1.5 GeV. The last column comprises the total
finite GDH integral from $\pi$, $\eta$, and $\pi\pi$ photoproduction and 
in case of the deuteron from photodisintegration.} 
\begin{ruledtabular}
\begin{tabular}{c|dddddd}
 & \multicolumn{1}{c}{$\pi^+\pi^-$} & \multicolumn{1}{c}{$\pi^0\pi^0$} & 
\multicolumn{1}{c}{$\pi^+\pi^0$} & \multicolumn{1}{c}{$\pi^-\pi^0$} &
\multicolumn{1}{c}{$\Sigma \,\pi\pi$} & \multicolumn{1}{c}{total [$\mu$b]} \\
\colrule   
neutron   & 22.23 & 1.82 & 33.28 &   & 57.33 & 190.51\\
proton    & 16.36 & 1.60 &      & 31.01 & 48.97 & 216.58 \\
deuteron   & 43.16 & 3.53 & 29.46 & 27.98 & 104.13 & -27.90 \\
\end{tabular}
\end{ruledtabular}
\end{table}
%%%%%%%%%%%%%%%%%%%%%%%%%%%%%%%%%%%%%%%%%%%%%%%%%%%%%%%%%%%%%%%%%%%%%%%%%

%% file: pap3_fig.tex
\begin{figure}
\includegraphics[scale=.8]{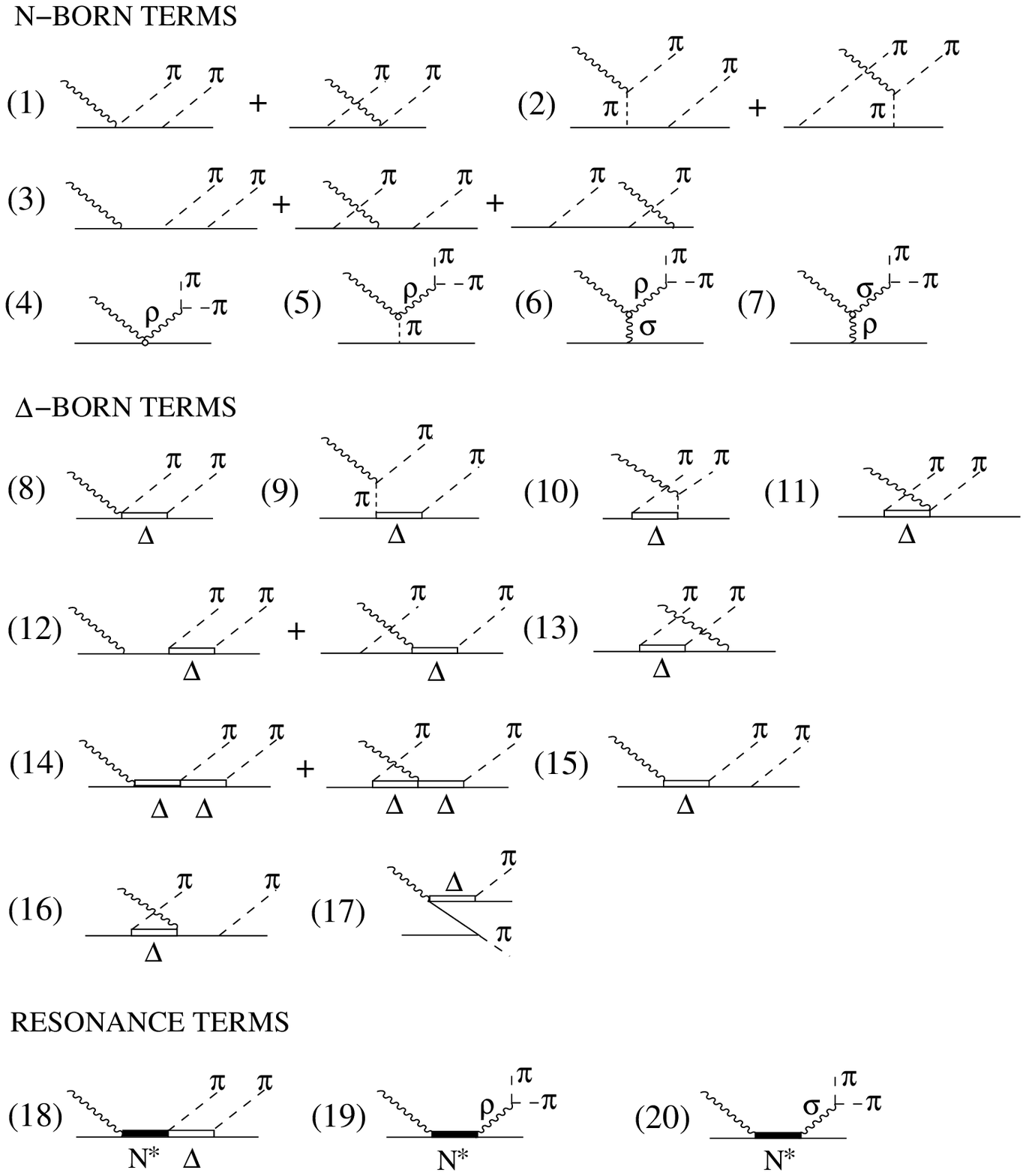}
\caption{Diagrams for the reaction $\gamma N\to\pi\pi N$ used in
the present work.} 
\label{fig1}
\end{figure}

\begin{figure}
\includegraphics[scale=.8]{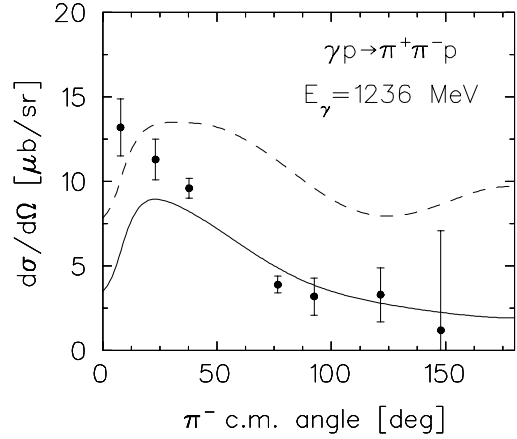}
\caption{$\pi^-$ angular distribution for $\gamma
N\to\pi^-\pi^+ p$. The solid (dashed) curves are calculated with
(without) absorption correction. The data are taken from 
Ref.~\protect\cite{Hauser}.}
\label{fig3}
\end{figure}

\begin{figure}
\includegraphics[scale=.7]{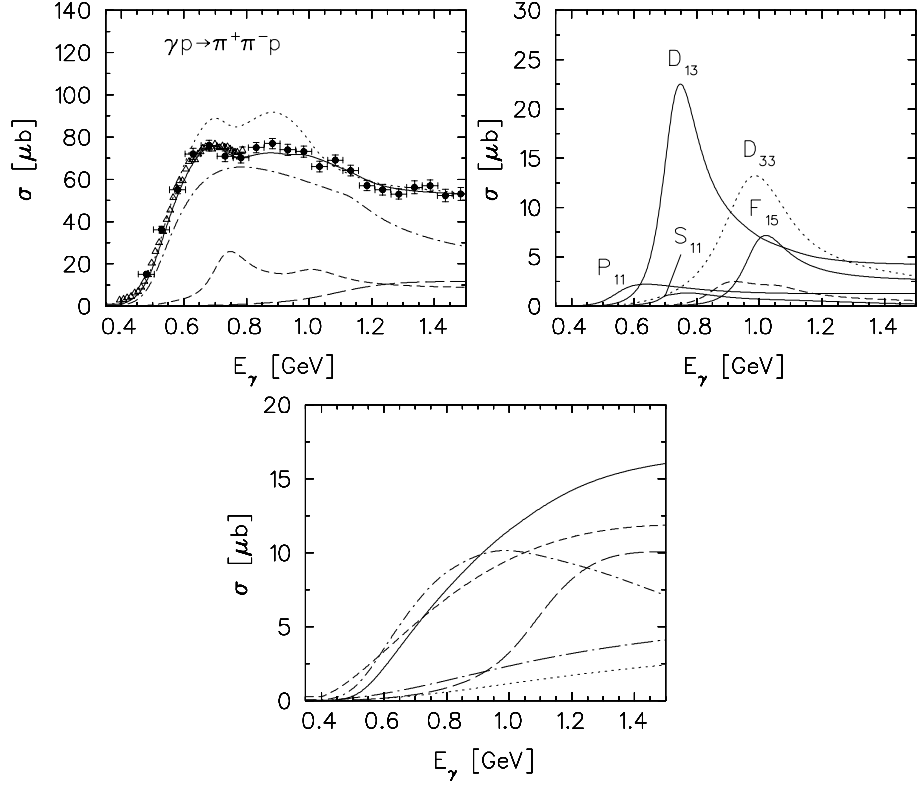}
\caption{Total cross section for $\gamma p\to\pi^+\pi^- p$. 
Contributions of individual diagrams as follows: Upper left
panel: dash-dotted: 
$\Delta$ Kroll-Ruderman term plus pion-pole term (diagrams (8) and (9)); 
short dashed: 
all resonance terms (diagrams (18),(19) and (20)); 
long dashed: $\rho^0$-photoproduction via $\pi^0$ 
and $\sigma$-exchange (diagrams (5) and (6)); 
solid: resulting cross section without $D_{33}(1700)$; dotted: additional 
inclusion of $D_{33}(1700)$.
The data are from Ref.~\protect\cite{ABBHHM} (circles) and 
Ref.~\protect\cite{Bragh} (triangles).
Upper right panel: solid: contributions 
of $P_{11}(1440)$, $D_{13}(1520)$, $F_{15}(1680)$, and  $S_{11}(1535)$; 
dotted: contribution of $D_{33}(1700)$; 
dashed: combined contribution of $S_{31}(1620)$, $P_{13}(1720)$, and 
$D_{15}(1675)$.
Lower panel: solid: contribution of diagrams (12); short-dashed: 
diagrams (1) and (2); long-dashed: diagram (6); 
short dash-dotted: diagrams (14); dotted: diagrams (3); 
long-dash-dotted: remaining Born terms. 
}
\label{fig2}
\end{figure}

\begin{figure}
\includegraphics[scale=.8]{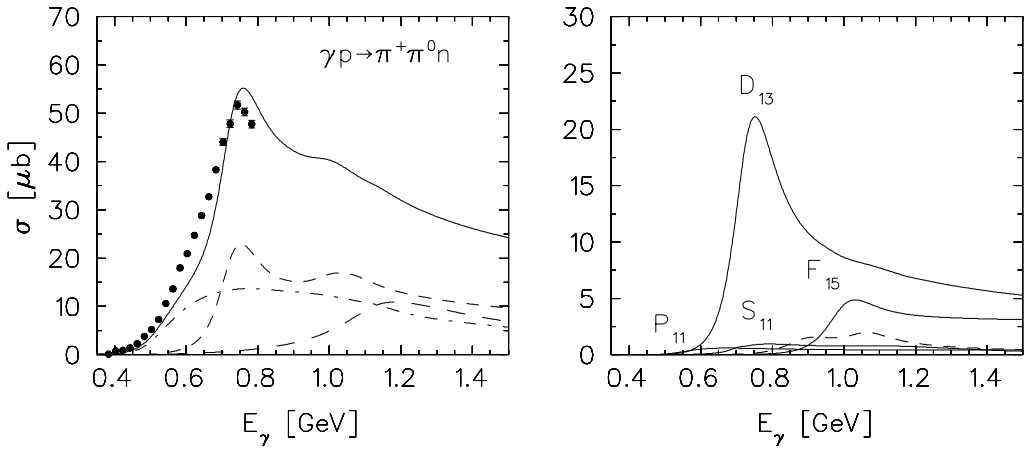}
\caption{Same as in Fig.~\protect\ref{fig2} for 
$\gamma p\to\pi^+\pi^0 n$. Long-dashed line on the left panel is the 
contribution of the nonresonant $\rho^+$-photoproduction (contact term (4) and 
$\pi$-exchange term (5)). 
The data are from Ref.~\protect\cite{Lang}.} 
\label{fig2a}
\end{figure}

\begin{figure}
\includegraphics[scale=.8]{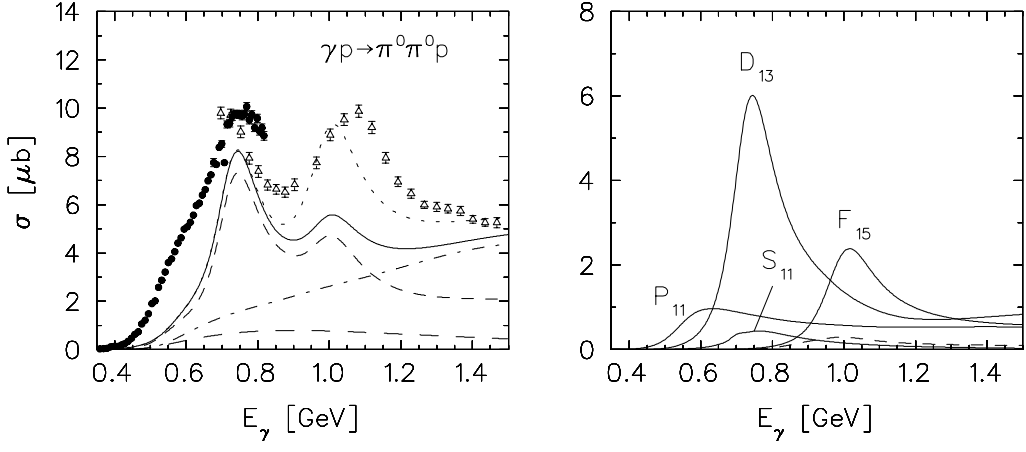}
\caption{Total cross section for
$\gamma p\to\pi^0\pi^0 p$. Left panel: dash-dotted: 
contribution of $N\Delta$ $s$- and $u$-channels (diagrams (12)); 
long-dashed: contribution of the $Z$-graph (diagram (17));  
dotted: calculation with positive sign of the 
$F_{15}(1680)\to\pi\Delta$ amplitude as predicted in \protect\cite{Capstic}. 
Experimental data from Ref.~\protect\cite{Wolf} (circles) and 
Ref.~\protect\cite{GRAAL} (triangles).
Right panel: solid: contributions 
of $P_{11}(1440)$, $D_{13}(1520)$, $F_{15}(1680)$, and  $S_{11}(1535)$; 
dotted: contribution of $D_{33}(1700)$; 
dashed: combined contribution of $S_{31}(1620)$, $P_{13}(1720)$, and 
$D_{15}(1675)$.
} 
\label{fig2b}
\end{figure}

\begin{figure}
\includegraphics[scale=.8]{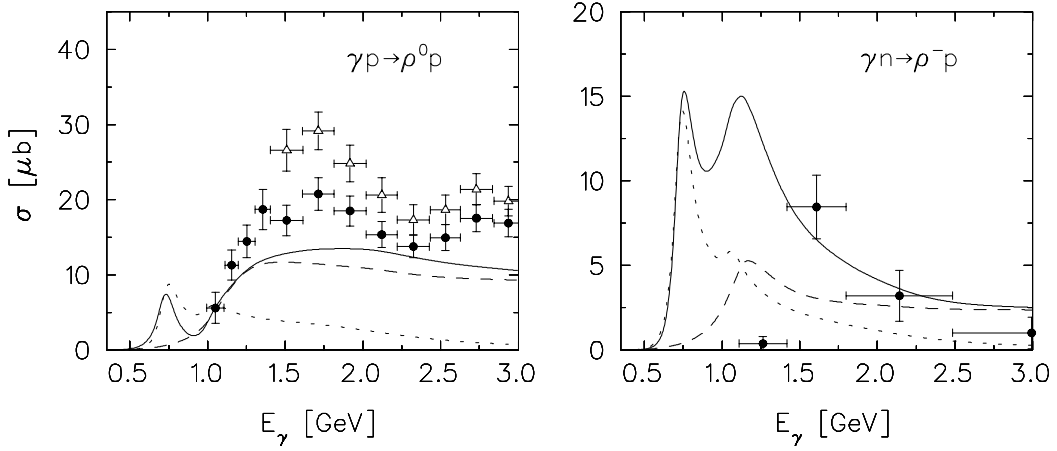}
\caption{Total cross sections for $\rho$ photoproduction.
Dotted curves: contribution from baryonic resonances in the 
$s$-channel. Left panel: dashed curve: contribution of $\sigma$-exchange 
(diagram (6)). Experimental data from Ref.~\protect\cite{ABBHHM}.
Right panel: dashed curve: contact term (4). Experimental data from 
Ref.~\protect\cite{ABHHM1}.} 
\label{fig2rho}
\end{figure}

\begin{figure}
\includegraphics[scale=.65]{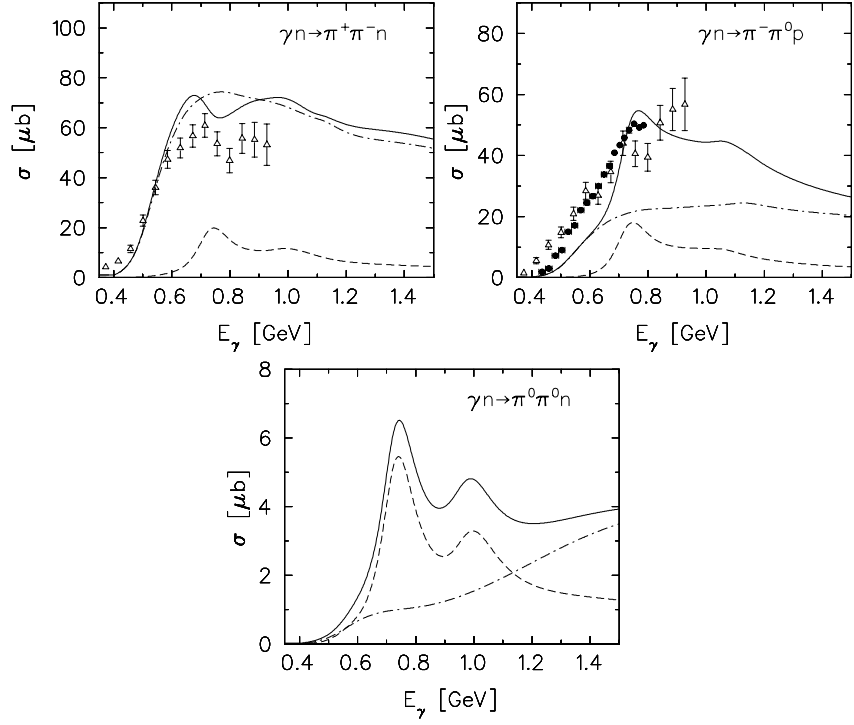}
\caption{Total cross section for double pion photoproduction on 
the neutron for different charge channels. Dashed curves: only 
resonance diagrams; dash-dotted curves: Born diagrams alone. Experimental 
data from Ref.~\protect\cite{Carbonara} (triangles) and 
Ref.~\protect\cite{Zabrod} (circles).}
\label{fig6}
\end{figure}

\begin{figure}
\includegraphics[scale=.65]{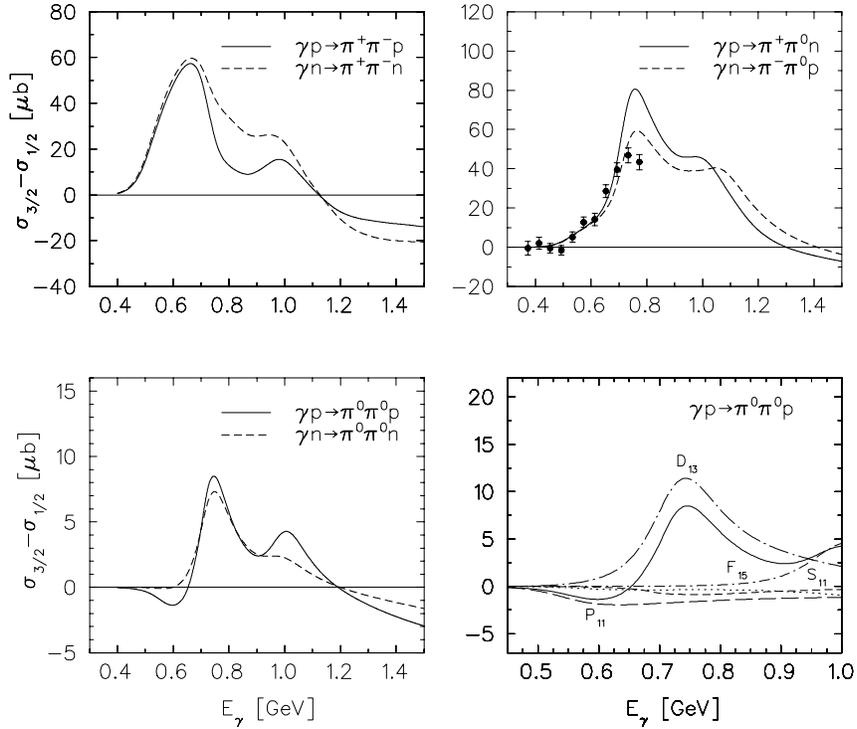}
\caption{Helicity asymmetry $\sigma_{3/2}-\sigma_{1/2}$ 
(\protect\ref{50}) for $\gamma N\to\pi\pi N$ in different charge channels. 
Lower right panel: individual contributions of 
$P_{11}(1440)$, $D_{13}(1520)$, $S_{11}(1535)$, and $F_{15}(1680)$
resonances. Dotted curve represents the combined 
contribution of Born diagrams and the solid curve the total 
asymmetry.}
\label{fig5}
\end{figure}

\begin{figure}
\includegraphics[scale=.6]{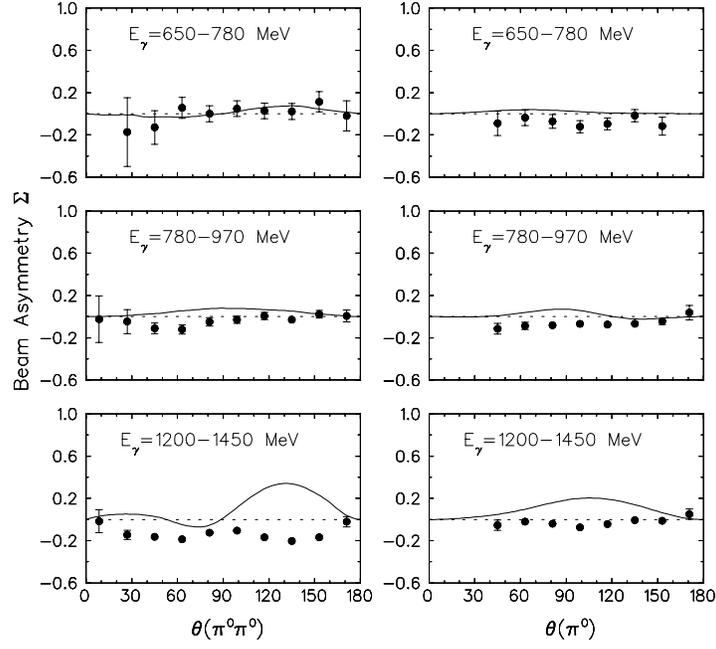}
\caption{Beam asymmetry $\Sigma$ for linearly polarized photons 
for $\gamma p\to\pi^0\pi^0p$. On the left panels $\theta(\pi^0\pi^0)$
denotes the angle of the total two-pion momentum, and on the right panels
$\theta(\pi^0)$ refers to the angle of one of the produced pions.
The data are from Ref.~\protect\cite{GRAAL}.}
\label{fig5a}
\end{figure}

\begin{figure}
\includegraphics[scale=.65]{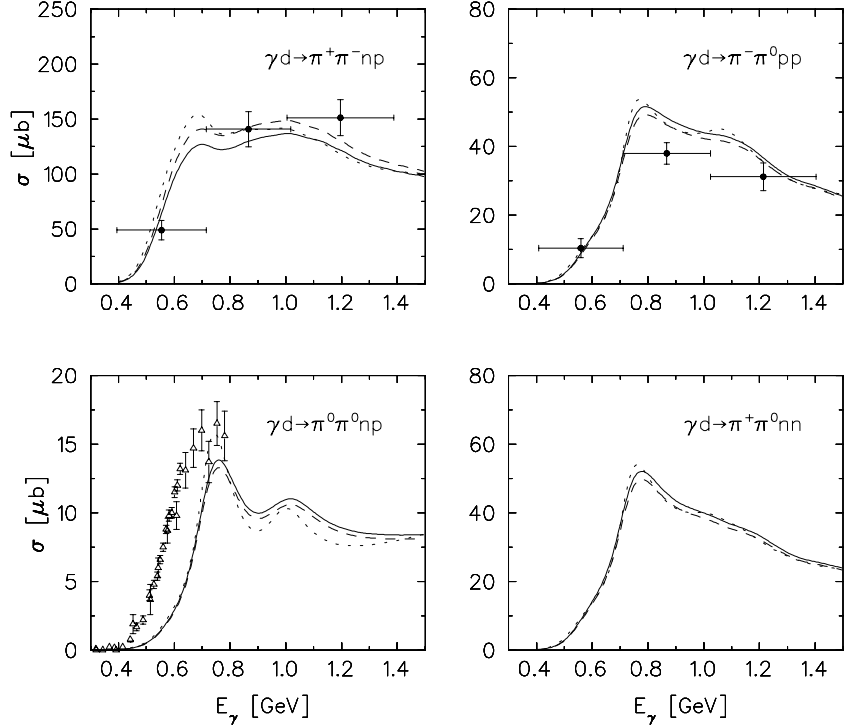}
\caption{Total cross section for incoherent double pion
photoproduction on a deuteron for different charge channels. Solid and 
dashed curves are obtained with and
without final $NN$ interaction. Dotted curves show the 
corresponding elementary cross sections. In $\pi^+\pi^-$ and $\pi^0\pi^0$ 
channels they are calculated as a sum of the cross sections on a proton and a 
neutron. The data are from 
Ref.~\protect\cite{ABHHM2} (circles) and Ref.~\protect\cite{Kru} 
(triangles).}
\label{fig7}
\end{figure} 

\begin{figure}
\includegraphics[scale=.8]{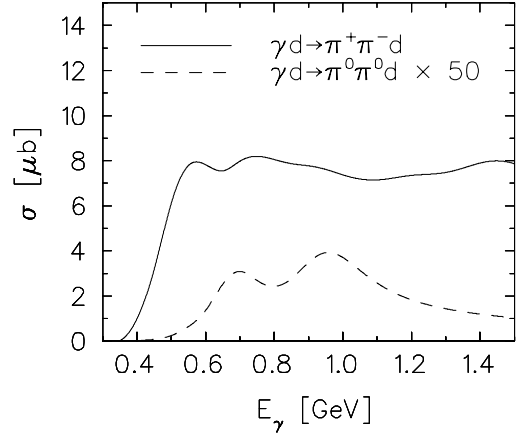}
\caption{Total cross sections for coherent double pion
photoproduction on a deuteron. The solid and the dashed curves 
represent the $\pi^+\pi^-$ and $\pi^0\pi^0$ channels, respectively. 
The $\pi^0\pi^0$ cross section is multiplied by a factor 50.}
\label{fig8}
\end{figure} 

\begin{figure}
\includegraphics[scale=.7]{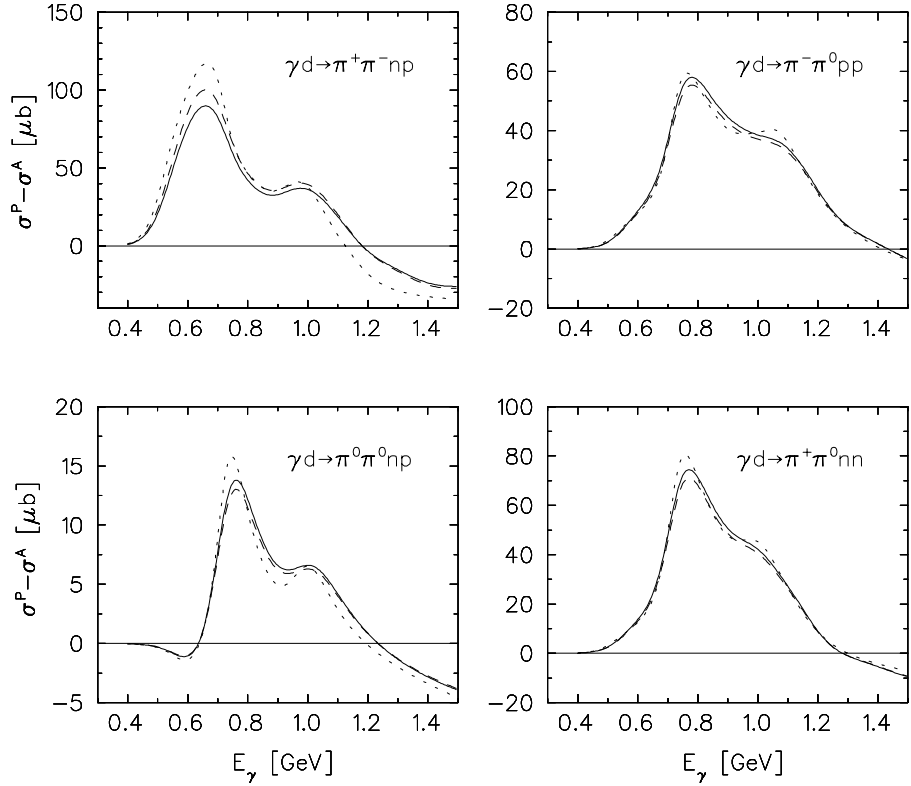}
\caption{Helicity asymmetry for incoherent double pion
photoproduction on a deuteron. The notation of the curves is as in 
Fig.~\protect\ref{fig7}.}
\label{fig7a}
\end{figure}